\documentclass[aps,prb,showpacs,amsmath,amssymb,reprint,10pt,superscriptaddress,longbibliography]{revtex4-2}
\usepackage[utf8]{inputenc}
\usepackage[colorlinks=true,urlcolor=blue,citecolor=red]{hyperref}
\usepackage{url}

\usepackage{siunitx}
\usepackage{physics}
\usepackage{amsthm}
\usepackage{graphicx}
\usepackage{enumerate}
\usepackage{dsfont}

\begin{document}
    
\title{A cavity quantum electrodynamics implementation of the Sachdev--Ye--Kitaev model}

\author{Philipp Uhrich}
\thanks{These two authors contributed equally.}
\affiliation{Pitaevskii BEC Center, CNR-INO and Dipartimento di Fisica, Università di Trento, Via Sommarive 14, Trento, I-38123, Italy}
\affiliation{INFN-TIFPA, Trento Institute for Fundamental Physics and Applications, Via Sommarive 14, Trento, I-38123, Italy}

\author{Soumik Bandyopadhyay}
\thanks{These two authors contributed equally.}
\affiliation{Pitaevskii BEC Center, CNR-INO and Dipartimento di Fisica, Università di Trento, Via Sommarive 14, Trento, I-38123, Italy}
\affiliation{INFN-TIFPA, Trento Institute for Fundamental Physics and Applications, Via Sommarive 14, Trento, I-38123, Italy}

\author{Nick Sauerwein}
\affiliation{Institute of Physics and Center for Quantum Science and Engineering, Ecole Polytechnique Fédérale de Lausanne (EPFL), Route cantonale, Lausanne, 1015, Switzerland}

\author{Julian Sonner}
\affiliation{Department of Theoretical Physics, University of Geneva, 24 quai Ernest-Ansermet, Gen\`{e}ve 4, 1211, Switzerland}

\author{Jean-Philippe Brantut}
\affiliation{Institute of Physics and Center for Quantum Science and Engineering, Ecole Polytechnique Fédérale de Lausanne (EPFL), Route cantonale, Lausanne, 1015, Switzerland}

\author{Philipp Hauke}
\email{philipp.hauke@unitn.it}
\affiliation{Pitaevskii BEC Center, CNR-INO and Dipartimento di Fisica, Università di Trento, Via Sommarive 14, Trento, I-38123, Italy}
\affiliation{INFN-TIFPA, Trento Institute for Fundamental Physics and Applications, Via Sommarive 14, Trento, I-38123, Italy}

\date{\today}

\begin{abstract}
The search for a quantum theory of gravity has led to the discovery of quantum many-body systems that are dual to gravitational models with quantum properties.
The perhaps most famous of these systems is the Sachdev-Ye-Kitaev (SYK) model. It features maximal scrambling of quantum information, and opens a potential inroad to experimentally investigating aspects of quantum gravity.
A scalable laboratory realisation of this model, however, remains outstanding.
Here, we propose a feasible implementation of the SYK model in cavity quantum electrodynamics platforms.
Through detailed analytical and numerical demonstrations, we show how driving a cloud of fermionic atoms trapped in a multi-mode optical cavity, and subjecting it to a spatially disordered AC-Stark shift retrieves the physics of the SYK model, with random all-to-all interactions and fast scrambling.
Our work provides a blueprint for realising the SYK model in a scalable system, with the prospect of studying holographic quantum matter in the laboratory.
\end{abstract}

\maketitle

\section{Introduction}
Holographic quantum matter is a postulated phase of strongly correlated many-body systems that hosts a variety of fascinating properties.
Chief among these are the existence of a holographically dual interpretation in terms of a quantum theory of gravity, and the attendant unconventional transport and chaotic characteristics \cite{HartnollLucasSachdev_Book, LiuSonner_2020}.
Despite the obvious wide-reaching interest across several physical disciplines from quantum matter to quantum gravity, such a holographic phase of matter has not to date been observed in any material or laboratory system.

One of the most paradigmatic of such systems, the Sachdev--Ye--Kitaev (SYK) model  \cite{SachdevYe_1993, Kitaev_talk}, was originally introduced in the study of strange metals and non-Fermi liquids \cite{Chowdhury_etal2022}, due to its lack of quasiparticle excitations resulting from an exponentially dense low-energy spectrum.
It describes $N$ fermions with interactions that are of infinite range, random, and uncorrelated. 
In addition to exhibiting rich strongly coupled many-body phenomena, the SYK model arguably hosts the simplest known example of holographic duality \cite{Sachdev_2015}. 
The model at large-$N$ and strong coupling shows an emergent SL$(2,\mathbb{R})$ symmetry \cite{parcollet1999non}, a maximal Lyapunov exponent \cite{Kitaev_talk}, as well as a sector of modes governed by the Schwarzian action \cite{Maldacena:2016hyu,Jensen:2016pah}, a set of properties it shares with two-dimensional Jackiw--Teitelboim gravity \cite{Teitelboim_1983, Jackiw_1985}.
The diversity of intriguing properties, in particular the exciting perspective of performing laboratory experiments on holographic systems, makes it highly desirable to find an experimental realization of this model.

In spite of its connections with strange metals in condensed-matter physics \cite{Chowdhury_etal2022}, no natural material is known that can microscopically realise the particular SYK interaction.
The search has therefore focused on artificial systems in solid-state mesoscopic systems \cite{PikulinFranz_2017, Chew_etal2017, Chen_etal2018}, cold atoms in optical lattices \cite{Danshita_etal2017, WeiSedrakyan_2021}, or on direct digital quantum simulation \cite{GarciaAlvarez_etal2017, Babbush_etal2019}.
Small-scale, minimal versions of the latter have being reported using nuclear magnetic resonance \cite{Luo_etal2019} and superconducting qubits \cite{Jaferris_etal2022}.
However, the $[N(N-1)/2]^2$ scaling of the number of independent couplings and their infinite range nature represents a formidable challenge for bottom-up approaches.
Reduced versions, such as the one implemented in Ref.~\cite{Jaferris_etal2022} appear to only partially retain the SYK physics \cite{Kobrin_etal2023}, highlighting the need for a concept realizing the dense set of couplings prescribed by the original model.
The experimental realization of the SYK model, specifically in the large-$N$ limit, thus remains an exciting challenge.

In this work, we propose an experimentally feasible quantum simulation of the SYK model, leveraging on recent advances in cavity quantum electrodynamics (cQED) architectures. 
This platform natively realises long-range all-to-all interactions \cite{Mivehvar_etal2021}.
This feature was used in proposals for the study of glassy physics \cite{StrackSachdev_2011,Mueller_etal2012,Buchhold_etal2013}, and it has been used recently to implement hyperbolic interactions, relevant to the study of holographic quantum matter \cite{Periwal_etal2021}.
Furthermore, controlled disorder in the light-matter coupling \cite{Sauerwein_etal2022,Baghdad_etal2022} as well as degenerate fermionic atoms \cite{Roux_etal2020,Zhang_etal2021} are now available in the cQED framework.
As we show, this combination, together with the multimode nature of optical cavities, allows for the realization of the long-range, all-to-all and random SYK interaction in existing state-of-the art experimental systems with up to hundreds of particles.
Starting from a complete model of trapped Fermions in a high-finesse cavity, we identify two key physical parameters allowing for the dynamics of the effective model to replicate that of the SYK model: 
(i) the effective number of modes participating in the light-matter interactions, which can be tuned by varying the spacing of the cavity modes with respect to the drive--cavity detuning, allowing one to reach the chaotic regime \cite{LantagneHurtubise_etal2018,Kim_etal2020}, and (ii) the ratio of cavity mode waist to the harmonic length of the dipole trap, controlling the mechanical coupling between atoms and cavity photons.

Further, we perform first-principles numeric calculation of the interaction amplitudes over a broad range of experimentally available parameters. Unexpectedly, we uncover a feature common to existing proposals \cite{Chen_etal2018, WeiSedrakyan_2021}, but unnoticed so far; namely that the statistical distribution of the interaction amplitudes is not Gaussian, but interpolates between a Gaussian and a Cauchy distribution.
Despite this deviation from the original SYK prescription, the dynamics generated by these interaction show good agreement with those of the ideal SYK model.
In particular, we numerically exactly simulate out-of-time-order correlators and the spectral form factor of the effective model over a range of system parameters, showing that both quantities approach their SYK counterparts as the effective number of cavity modes is increased.
This work shows that the SYK model is within the reach of cQED-based experiments, and it sheds further light on the robustness of SYK physics against experimentally motivated imperfections.

\begin{figure}[t!]
	\centering
    \includegraphics[width=\linewidth]{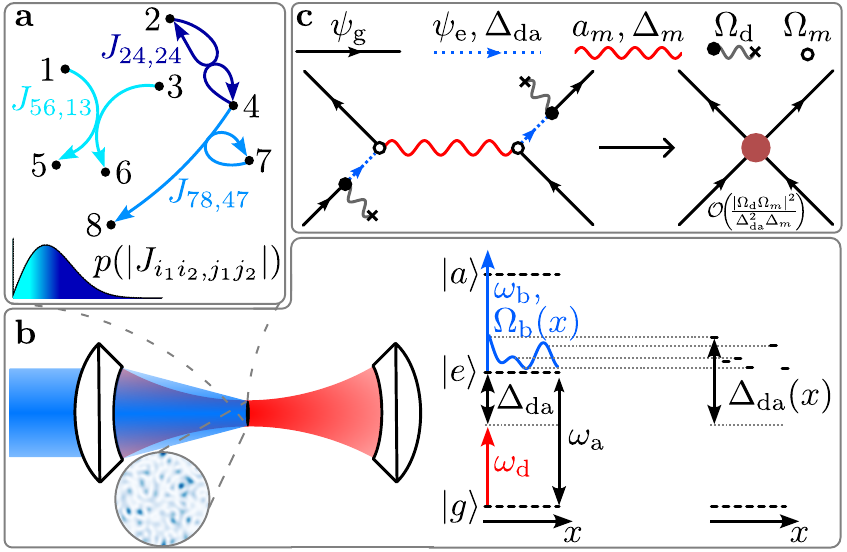}
	\caption{\textbf{Realizing the SYK model in a multi-mode cavity.}
		(a) Representation of the target model $ H_\mathrm{SYK} $ [Eq.~\eqref{e:Hsyk}] as a directed hypergraph for $N=8$ modes.
        Each hyperedge (only three are shown) corresponds to a complex interaction amplitude $ J_{i_1 i_2 ; j_1 j_2} $ with $i_1>i_2,j_1>j_2$, whose magnitude (shades of blue) follows a Rayleigh distribution $p(\abs{ J_{i_1 i_2 ; j_1 j_2} })$.
        (b) Cavity-QED setup (left):
        A quasi-two-dimensional cloud of fermionic atoms (black ellipse) is trapped at an antinode of a longitudinal mode (red) of a multi-mode optical cavity.
        A random phase-mask is imprinted on a light-shift beam (blue) which is focused into the atomic cloud via a lens attached to the cavity mirror, and thus creates a disordered intensity distribution (speckle, lower circle).
        The required atomic level structure (centre) consists of a ground $\ket{g}$, excited $\ket{e}$, and auxiliary $\ket{a}$ electronic state for an atom at position $x$ (horizontal black arrow).
        The drive beam is far red detuned by $\abs{ \Delta_{\mathrm{da}} }$ from the $g$--$e$ transition at frequency $\omega_\mathrm{a}$, allowing for adiabatic elimination of state $\ket{e}$.
        The light-shift beam (blue), with speckled intensity $\Omega_{\mathrm{b}}(x)$, off-resonantly couples states $\ket{a}$ and $\ket{e}$, thereby inducing a position-dependent AC-Stark shift of the excited state energy (right).
        The cavity-mediated interactions are randomised via the disordered drive--atom detuning $ \Delta_{\mathrm{da}} (x)$.
        (c) Feynman diagram for the fourth-order process (left) which yields the long-ranged all-to-all interactions $ \mathcal{J}_{i_1 i_2 ; j_1 j_2}  \sim \mathcal{O}(\abs{\Omega_\mathrm{d} \Omega_m}^2 /  \Delta_{\mathrm{da}} ^2  \Delta_m )$ of the effective model $H_\mathrm{eff}$ (right, large greyish-red vertex) after adiabatic elimination of the excited-state (blue dashed edges), and integrating-out the cavity modes (red wavy edge).
        On the left,  grey wavy edges ending in a cross represent the classical drive field, and solid(empty) vertices indicate drive--atom(cavity--atom) interactions.
	}
	\label{f:concept}
\end{figure}

\section{Results}
Our target is the SYK model with complex two-body interactions among $N$ spinless Dirac fermions \cite{Sachdev_2015}, a variant of the $q=4$ Majorana model \cite{Maldacena:2016hyu}, described by the Hamiltonian
\begin{equation}\label{e:Hsyk}
     H_\mathrm{SYK}  = \frac{1}{(2N)^{3/2}}\sum_{i_1, i_2, j_1, j_2 = 1}^N  J_{i_1 i_2 ; j_1 j_2}  c^\dagger_{i_1} c^\dagger_{i_2} c_{j_1} c_{j_2} .
\end{equation}
Here, $c^\dagger_i$ and $c_i$ satisfy the canonical anticommutation relations, and respectively denote the creation and annihilation operator for the $i$th fermion mode. 
The model falls within a broader class of two-body random ensembles and embedded Gaussian unitary matrices, which have long been studied in the context of many-body quantum chaos and nuclear shell models \cite{Brody_etal1981}.
In the target model, the interaction amplitudes $J_{i_1 i_2 ; j_1 j_2} = J_{j_1 j_2 ; i_1 i_2}^*$ are complex random variables, with real and imaginary parts sampled independently and identically from normal distributions, which have variances $\mathrm{var} \! \left( \mathrm{Re} J_{i_1 i_2 ; j_1 j_2} \right) = J^2$, $\mathrm{var} \! \left( \mathrm{Im} J_{i_1 i_2 ; j_1 j_2} \right) = 0$  if $i_1=j_1, i_2=j_2$, and $\mathrm{var} \! \left( \mathrm{Re} J_{i_1 i_2 ; j_1 j_2} \right) = \mathrm{var} \! \left( \mathrm{Im} J_{i_1 i_2 ; j_1 j_2} \right) = J^2/2$ otherwise.
The model can be represented as a directed hypergraph on $N$ nodes with $[N(N-1)/2]^2$ hyperedges that have complex weights in $\lbrace  J_{i_1 i_2 ; j_1 j_2}  \vert i_1>i_2, j_1>j_2 \text{ and } i_k,j_k = 1,\ldots,N \text{ for } k=1,2 \rbrace$, and are oriented from nodes $j_1$, $j_2$ to nodes $i_1$, $i_2$, see Fig.~\ref{f:concept}a.
Physically, the model is zero-dimensional due to its all-to-all connectivity.
Realizing these all-to-all interactions in a random and uncorrelated way is a formidable experimental challenge.

\subsection{Experimental approach}
We describe the general approach in what follows, and defer specific experimental numbers, in particular for the platform of Ref.~\cite{Sauerwein_etal2022} using  $^6$Li  atoms, to the end of this article. 
Our envisaged cQED setup is sketched in Fig.~\ref{f:concept}b (left). 
It consists of a Fermi gas trapped in a two-dimensional (pancake) geometry at the antinode of a longitudinal cavity mode.
We consider the trap as harmonic and use the motional eigenstates in the plane to represent the modes of $ H_\mathrm{SYK} $.
The mechanical coupling between an atom and a cavity photon is parameterised by the ratio $\zeta \equiv  x_0  / (w_0 /\sqrt{2})$, where  $ x_0 $ is the oscillator length associated to the trapping potential, and $w_0$ is the waist of the fundamental cavity mode.

A pump laser, with Rabi(angular) frequency $\Omega_\mathrm{d}$($\omega_\mathrm{d}$) drives a transition between a ground, $\ket{g}$, and excited, $\ket{e}$, electronic level, see Fig.~\ref{f:concept}b (centre), either from the side or along the cavity axis. 
We denote the associated transition frequency as $\omega_\mathrm{a}$, and the detuning of the drive therefrom as $ \Delta_{\mathrm{da}}  \equiv \omega_\mathrm{d} - \omega_\mathrm{a}$. For sufficiently large $ \abs{\Delta_{\mathrm{da}} }$, the excited state can be adiabatically eliminated, such that the $N$ atoms encode $N$ complex, spinless fermions.
We make a long-wavelength approximation for the amplitude of the drive beam $g_\mathrm{d}(\boldsymbol{r})=1$ over the spatial extent of the atomic cloud, which supposes either on-axis pumping or a very-low angle from the side.
Simulations illustrating the qualitative differences for drive amplitudes with non-uniform phases are reported in the Fig.~\ref{f:convergence_test}h~and~i.

The cavity mediates long-ranged fermion--fermion interactions thanks to the virtual exchange of photons between atoms at arbitrary positions $\boldsymbol{r}$ within the pancake, yielding a fully connected, zero dimensional geometry. 
Approaching the SYK model of Eq.~\eqref{e:Hsyk} requires to render these random and independent between different fermion mode $4$-tuples.

To realise two-body interactions with random amplitudes $ J_{i_1 i_2 ; j_1 j_2} $, we propose to dress the excited state $\ket{e}$ with light near-resonant with a transition to a higher excited state.
Using a random intensity distribution such as a speckle pattern for the dressing produces a random light shift, proportional to the local intensity.
The drive--atom detuning $ \Delta_{\mathrm{da}} (\boldsymbol{r})$, and thus the photon-mediated interaction, inherits this random character, similar to Ref.~\cite{Sauerwein_etal2022}. 

This scheme however only leads to a separable (low-rank) SYK-type model, where the amplitudes $ J_{i_1 i_2 ; j_1 j_2} $ are still correlated. It was shown in Ref.~\cite{Kim_etal2020} that coupling to a number of modes which is (super)extensive in the system size $N$ reduces the correlations and allows for the manifestation of the SYK physics, such as quantum chaos. As we show below, this naturally occurs thanks to the multimode structure of optical Fabry-Perot cavities, even in cases where the cavity modes are not exactly degenerate \cite{Vaidya_etal2018}.

\subsection{Microscopic Hamiltonian}
The many-body Hamiltonian realised by the setup described above is given, within a reference frame rotating at the drive frequency $\omega_\mathrm{d}$, by 
\begin{equation}
     H_\mathrm{mb}=  H_{\mathrm{kt}}  + H_\mathrm{c} + H_\mathrm{a} +  H_\mathrm{ac}  +  H_\mathrm{ad}  \,.
\end{equation}
Here, 
\begin{equation}
     H_{\mathrm{kt}}  = \sum_{\mathrm{s}=\mathrm{e},\mathrm{g}} \int d^2r \psi_\mathrm{s}^\dagger(\boldsymbol{r}) \left( \frac{\boldsymbol{p}^2}{2  m_{\mathrm{at}}  } +   V_{\mathrm{t}} (\boldsymbol{r}) \right) \psi_\mathrm{s}(\boldsymbol{r}) \,,
\end{equation}
governs the dynamics of the atomic centre-of-mass (external) degrees of freedom, where $ m_{\mathrm{at}} $ is the atomic mass, $ V_{\mathrm{t}} (\boldsymbol{r})$ is the harmonic trapping potential, and $\psi_\mathrm{g}(\boldsymbol{r})$[$\psi_\mathrm{e}(\boldsymbol{r})$] is the field operator for the ground(excited) electronic state. 
The integral runs over the plane of the atomic cloud.
The term $H_\mathrm{c} = \sum_m  \Delta_m  a^\dagger_m a_m $ represents the total energy of  multiple cavity modes, with integer index $m\geq0$, detunings $ \Delta_m  = \omega_m - \omega_\mathrm{d}$ (we set $\hbar=1$ throughout), and corresponding photonic creation(annihilation) operators $a^\dagger_m$($a_m$).
The contribution ${H_\mathrm{a} = -\int d^2r  \Delta_{\mathrm{da}}  \psi_\mathrm{e}^\dagger (\boldsymbol{r}) \psi_\mathrm{e} (\boldsymbol{r})}$ represents the excited-state energy.
The atom--cavity and atom--drive interactions are (within the rotating wave approximation) respectively given by
\begin{align}
     H_\mathrm{ac}  &= \frac{1}{2} \sum_m \int d^2r \left( \Omega_m g_m(\boldsymbol{r}) \psi_\mathrm{e}^\dagger(\boldsymbol{r}) \psi_\mathrm{g}(\boldsymbol{r}) a_m + \mathrm{H.c.} \right) , \\
     H_\mathrm{ad}  &= \Omega_\mathrm{d} \int d^2r \left( g_\mathrm{d}(\boldsymbol{r}) \psi_\mathrm{e}^\dagger(\boldsymbol{r}) \psi_\mathrm{g}(\boldsymbol{r})  + \mathrm{H.c.} \right),
\end{align}
where $g_m(\boldsymbol{r})$[$g_\mathrm{d}(\boldsymbol{r})$] are the amplitudes of the cavity(drive) modes, and $\Omega_m$ is the atom--cavity coupling strength of the $m$th cavity mode.
Concretely, we will consider multiple transverse cavity modes (TCMs), whose amplitudes $g_m(\boldsymbol{r})$ over the plane of the atomic cloud are given by two-dimensional Hermite--Gauss modes.
The latter are denoted by two transverse-mode indices $n_x,n_y$, which we map to a unique integer index $m\geq 0$ [see Fig.~\ref{f:freqs}, and below Eq.~\eqref{e:sm_Hmb}], and whose sum we denote as $m_\Sigma$. 
The TCM frequencies can then be written as $\omega_m = \omega_\mathrm{c} + m_\Sigma \delta \omega$, where $\omega_\mathrm{c} \equiv \omega_{m=0}$ and $\delta \omega$ is the frequency spacing of the TCMs.

With these, we have all the ingredients to derive an effective Hamiltonian $H_\mathrm{eff}$ describing SYK-type physics. 
(Since $H_\mathrm{eff}$ is composed of spin-polarised fermions, Pauli exclusion prevents contact interactions, so we ignore them in $H_\mathrm{mb}$ \emph{ab initio} \cite{Giorgini_etal2008}.)
Formally, we introduce disorder into the system via a random spatial modulation of the atomic resonance $\omega_\mathrm{a}(\boldsymbol{r})$, equivalently of the drive--atom detuning $ \Delta_{\mathrm{da}} (\boldsymbol{r})$.
This spatial disorder propagates into the two-body interactions of our effective model, thereby randomizing them.
We now sketch the main steps for the derivation of our effective model, with further details given in the Methods.

\subsection{Effective model}
First, we assume $\abs{  \Delta_{\mathrm{da}} (\boldsymbol{r}) }$ to be the dominant energy scale at all positions $\boldsymbol{r}$, such that we may adiabatically eliminate the excited state.
We thus replace $\psi_\mathrm{e}(\boldsymbol{r})$ in $H_\mathrm{mb}$ through
\begin{equation}\label{e:psi_e_adelim}
    \psi_\mathrm{e}(\boldsymbol{r}) = \left[\frac{\Omega_\mathrm{d} g_\mathrm{d}(\boldsymbol{r})
        }{ \Delta_{\mathrm{da}} (\boldsymbol{r}) } + \frac{1}{2} \sum_m \frac{\Omega_m g_m(\boldsymbol{r})
        }{ \Delta_{\mathrm{da}} (\boldsymbol{r}) } a_m \right]\psi_\mathrm{g}(\boldsymbol{r}) \,.
\end{equation}

Additionally, the relation $\abs{ \Omega_\mathrm{d} \Omega_m } / \abs{  \Delta_{\mathrm{da}}   \Delta_m  } \ll 1$ allows us to decouple the atomic and cavity degrees of freedom using a Schrieffer--Wolff transformation, giving the effective Hamiltonian $H_\mathrm{eff} = e^S H e^{-S}$. 
Truncating at second order, the transformation is generated by  
\begin{equation}
    S = -   \sum_m  \left( \frac{ 1 }{   \Delta_m } a_m \Theta_m - \mathrm{H.c.} \right) ,
\end{equation}
with $ \Theta_m =  \int d^2r \Omega_\mathrm{d}^* g_\mathrm{d}^*(\boldsymbol{r}) \Omega_m g_m(\boldsymbol{r}) \psi_\mathrm{g}^\dagger(\boldsymbol{r}) \psi_\mathrm{g}(\boldsymbol{r}) / (2  \Delta_{\mathrm{da}} (\boldsymbol{r}) ) $.

Finally, expanding the field operators in the eigenbasis that diagonalises the resulting single-body contribution, $\psi_\mathrm{g}(\boldsymbol{r}) = \sum_{i_1} \phi_{i_1}(\boldsymbol{r}) c_{i_1}$, we obtain the effective Hamiltonian in terms of spinless, complex fermions, as required by the target model in Eq.~\eqref{e:Hsyk},
\begin{equation}\label{e:Heff_main}
    H_\mathrm{eff} = \sum_{i_1} \epsilon_{i_1} c^\dagger_{i_1} c_{i_1} + \sum_{i_1, i_2, j_1, j_2} \mathcal{J}_{i_1 i_2 ; j_1 j_2} c^\dagger_{i_1} c^\dagger_{i_2} c_{j_1} c_{j_2} \,.
\end{equation}
The above procedure yields long-range all-to-all two-body interactions via a fourth order process, as shown diagrammatically in Fig.~\ref{f:concept}c, at an energy scale ${ \mathcal{E} = ( \abs{ \Omega_\mathrm{d}}^2 /  \Delta_{\mathrm{da}}  )( \abs{ \Omega_{m=0}}^2 /  \Delta_{\mathrm{da}}  ) /  \Delta_{\mathrm{cd}}  }$, where ${  \Delta_{\mathrm{cd}}  \equiv \Delta_{m=0} }$ (see also Methods Sec.~\ref{s:derivation_Heff}, and SI Sec.~\ref{s:energies}).
The antisymmetrised form of the corresponding interaction amplitudes are 
\begin{align}\label{e:Jmnpq_asym_main}
     \mathcal{J}_{i_1 i_2 ; j_1 j_2}  =  \mathcal{E} \sum_{m} \frac{  \Delta_{\mathrm{cd}} }{ \Delta_m } \bigl[  &\tilde{I}_{i_1 j_1, m}  \tilde{I}_{ j_2 i_2, m }^* - \tilde{I}_{i_2 j_1, m} \tilde{I}_{ j_2 i_1, m }^* \nonumber \\
    - & \tilde{I}_{i_1 j_2, m} \tilde{I}_{ j_1 i_2, m }^* + \tilde{I}_{i_2 j_2, m} \tilde{I}_{ j_1 i_1, m }^*  \bigr]   \,,
\end{align}
where the integrals
\begin{equation}\label{e:integral}
    \tilde{I}_{i_1 j_1, m} = \frac{1}{2} \int d^2r \frac{ g_\mathrm{d}(\boldsymbol{r}) g_m(\boldsymbol{r})^* \phi^*_{i_1}(\boldsymbol{r}) \phi_{j_1}(\boldsymbol{r}) }{ \Delta_{\mathrm{da}} (\boldsymbol{r}) /  \Delta_{\mathrm{da}} } 
\end{equation}
are randomised via the spatially disordered drive--atom detuning $ \Delta_{\mathrm{da}} (\boldsymbol{r})$.

The expression for $ \mathcal{J}_{i_1 i_2 ; j_1 j_2} $ shows that the contributions of the various cavity modes are suppressed with increasing mode index $m$ as $ \Delta_{\mathrm{cd}}  / \Delta_m  = 1/(1+m_\Sigma \delta \omega/ \Delta_{\mathrm{cd}} )$. 
In the limit of large ratio $\tilde{\delta \omega}=\delta \omega /  \Delta_{\mathrm{cd}} $, only the single mode $m=0$ contributes significantly to the dynamics, resulting in an antisymmetrised product  form of the amplitudes $\mathcal{J}_{i_1 i_2 ; j_1 j_2}$.
However, decreasing $\tilde{\delta \omega}$  increases the effective number of cavity modes mediating the two-body interactions, thus generating dynamics that approach those of the SYK model, as we show in the following section.

SYK models with one-body perturbations as in $H_\mathrm{eff}$ have been studied in, for instance, Ref.~\cite{GarciaGarcia_etal2018,Monteiro_etal21}, which showed that the chaoticity of the model is maintained for sufficiently weak perturbations.
In our case, the one-body part $\sum_{i_1} \epsilon_{i_1} c^\dagger_{i_1} c_{i_1}$ consists of $ H_{\mathrm{kt}} $, an effective dipole potential of order $\abs{\Omega_{\mathrm{d} }^2 /  \Delta_{\mathrm{da}}  }$, and a term stemming from normal ordering of the two-body part, of order $\mathcal{E}$ [see Eq.~\eqref{e:H_1body}--\eqref{e:Heff}]. 
The dipole potential can be compensated by introducing an additional dipole potential of equal magnitude and opposite sign at all $\boldsymbol{r}$.
For an example using the dipole potential induced by a laser driving directly the $\ket{g}$ to $\ket{a}$ transition, see the Supplementary Information (SI) \cite{supplementary}, Sec.~\ref{s:compensation}. 
The individual contributions to the normal-ordering term are energetically of the same magnitude as each $ \mathcal{J}_{i_1 i_2 ; j_1 j_2} $. However, its total strength scales as $N^{-2}$ relative to the desired two-body term, such that the latter rapidly dominates as $N$ is increased.
Finally, by increasing the drive power, $\mathcal{E}$ is enhanced as $\abs{\Omega_\mathrm{d}}^2$, permitting one to render $ H_{\mathrm{kt}} $ sufficiently weak relative to $ \mathcal{J}_{i_1 i_2 ; j_1 j_2} $.
In what follows, we therefore compare only the two-body (quartic fermion operator) term $ H_{\mathrm{eff}}^{(4)}  = \sum_{i_1, i_2, j_1, j_2}   \mathcal{J}_{i_1 i_2 ; j_1 j_2}  c^\dagger_{i_1} c^\dagger_{i_2} c_{j_1} c_{j_2}$ of $H_\mathrm{eff}$ to the full target Hamiltonian $ H_\mathrm{SYK} $.
To do so, we will normalise the interaction amplitudes of either model such that their ensemble variance is $J^2=1$.

\begin{figure}[t]
    \centering
    \includegraphics[width=\linewidth]{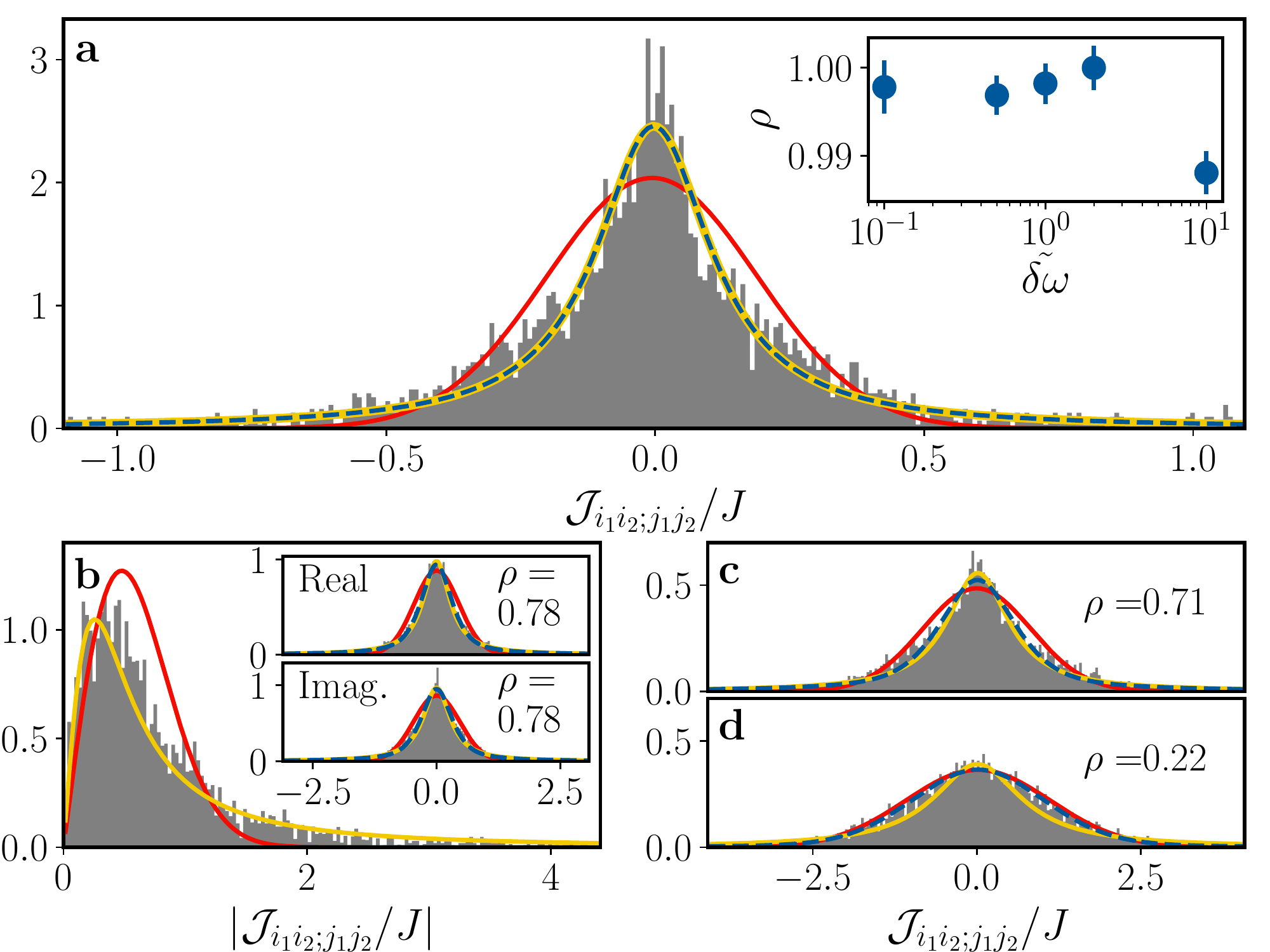}
    \caption{\textbf{Distribution of two-body interaction amplitudes}
        (a) Representative distribution of $\mathcal{J}_{i_1 i_2; j_1 j_2}$ from numeric calculation of Eq.~\eqref{e:Jmnpq_asym_main}, for $N=14$ fermionic modes, $\tilde{\delta \omega}=10^{-1}$, and $\zeta=1$. 
        Fitting the distribution by a (pseudo)Voigt profile \cite{Wertheim_etal1974} [blue dashed line, Eq.~\eqref{e:psv}], which interpolates with a parameter $\rho$ between a Gaussian ($\rho=0$, red line) and a Cauchy ($\rho=1$, yellow line) probability density, quantifies the deviation from Gaussian statistics.
        The interactions are more Cauchy than Gaussian for all $\tilde{\delta \omega}$ considered, as indicated by the inset, which shows the mean value (circles) and standard deviation (error-bars) of $\rho$ over $100$ independent disorder realizations.
        Numerical data of previous proposals shows similar features:
        (b) Data provided by the authors of a graphene-based proposal Ref.~\cite{Chen_etal2018}, is more Cauchy than Gaussian.
        (c), (d) Data provided by the authors of Ref.~\cite{WeiSedrakyan_2021}, show that depending on the hopping phase $\phi$ of their optical lattice proposal, the two-body interactions follow more closely a Cauchy [(c), $\phi=0$] or Gaussian distribution [(d), $\phi=\pi$].
    }
    \label{f:histograms}
\end{figure}

\subsection{Randomness of two-body interactions}
At this point, in previous works it is often assumed that for sufficiently many cavity modes the interaction amplitudes $\mathcal{J}_{i_1 i_2 ; j_1 j_2}$ as defined by Eq.~\eqref{e:Jmnpq_asym_main} will follow a Gaussian distribution, due to the central limit theorem \cite{Buchhold_etal2013,Danshita_etal2017,Mueller_etal2012}.
Here, instead, we numerically calculate the interactions $\mathcal{J}_{i_1 i_2 ; j_1 j_2}$ from first principles according to the microscopic expression of Eq.~\eqref{e:Jmnpq_asym_main}~and~\eqref{e:integral}.

First of all, we note that $ \mathcal{J}_{i_1 i_2 ; j_1 j_2}  \in \mathbb{R}$ because the eigenmodes $\phi_{k}(\boldsymbol{r})$ in Eq.~\eqref{e:integral} are real for all $k=i_1,i_2,j_1,j_2$, due to $H_\mathrm{eff}$ being time-reversal symmetric [see Eqs.\eqref{e:H_1body}--\eqref{e:Jmnpq_asym}].
The amplitudes' ensemble variance is then given by $J^2=\mathds{E}[ \mathcal{J}_{i_1 i_2 ; j_1 j_2} ^2] - \mathds{E}[ \mathcal{J}_{i_1 i_2 ; j_1 j_2} ]^2$, where $\mathds{E}[\cdot]$ denotes averaging over an ensemble of disorder realizations.
Figure~\ref{f:histograms}(a) shows the distribution for a single, representative, realization of $ H_{\mathrm{eff}}^{(4)}  / J$ (for details on the numeric calculation of the speckle and interaction amplitudes, see Methods, Sec.~\ref{s:numsim}).
The distribution deviates from a Gaussian probability density, and is better approximated by a Cauchy distribution.
Interestingly, this feature is not unique to the setup considered here: We have found that some previous proposals which also numerically calculate interactions according to the microscopic description of their respective model, have also obtained non-Gaussian distributions (see the data reproduced from Refs.~\cite{Chen_etal2018,WeiSedrakyan_2021} in Fig.~\ref{f:histograms}b~--~d), though the non-Gaussianity was not reported.  
A possible reason for the failure of the CLT may be the presence of correlations within each realization of $ H_{\mathrm{eff}}^{(4)} $, which could arise, for instance, between distinct interaction amplitudes that have one or more fermion modes in common and thus sample the spatial disorder in a similar way.
The important question then arises in how far the deviating probability distribution modifies the physics of the SYK model, a question that is currently actively researched also for other probability distributions \cite{Krajewski_etal2019,Cao_etal2020,GarciaGarcia_etal2021,Tezuka_etal2023}.  
In Sec.~\ref{s:cauchy_gauss} of the SI, we compare the dynamics of $ H_\mathrm{SYK} $ with Cauchy distributed interactions to its Gaussian counterpart, finding that they are qualitatively the same.

\subsection{Comparison to target model}
In this section, we numerically probe the chaoticity of the dynamics generated by $ H_{\mathrm{eff}}^{(4)} $, and compare them to the dynamics of the target model $ H_\mathrm{SYK} $ with real $ J_{i_1 i_2 ; j_1 j_2} $  \cite{Danshita_etal2017, Cao_etal2020}.
To this end, we simulate out-of-time-order correlators (OTOCs), and the spectral form factor (SFF) as probes of, respectively, early-time scrambling and late-time chaos \cite{Maldacena_etal2016, Kitaev_talk,  Kobrin_etal2021, Gharibyan_etal2018, Cotler_etal2017_1}.
We employ exact diagonalization methods, the limitation of which to small system sizes prevents one from accessing the maximal scrambling rate of the SYK model \cite{FuSachdev_2016}.
Thus, our goal is rather to use the dynamics of $ H_\mathrm{SYK} $ for a given accessible $N$ as a benchmark to which we can compare the dynamics of $ H_{\mathrm{eff}}^{(4)} $, and to show how the parameters entering $ H_{\mathrm{eff}}^{(4)} $ can be tuned such that its dynamics approach those of $ H_\mathrm{SYK} $.

Our simulations start by numerically calculating the interaction amplitudes $\mathcal{J}_{i_1 i_2 ; j_1 j_2}$ as defined by Eq.~\eqref{e:Jmnpq_asym_main}, with $m=0,1,\ldots,M$ (all data shown here is converged with respect to the cut-off $M$, see Fig.~\ref{f:convergence_test}b--e), for multiple independent disorder realizations.
For a given realization, we use the normalised amplitudes $ \mathcal{J}_{i_1 i_2 ; j_1 j_2} /J$ to construct $ H_{\mathrm{eff}}^{(4)} /J$, whose dynamics are then solved via exact diagonalization (see Methods, Sec.~\ref{s:numsim}).
The simulated time $t$ is thus in units of $J$.

\begin{figure}[h]
    \centering
    \includegraphics[width=\linewidth]{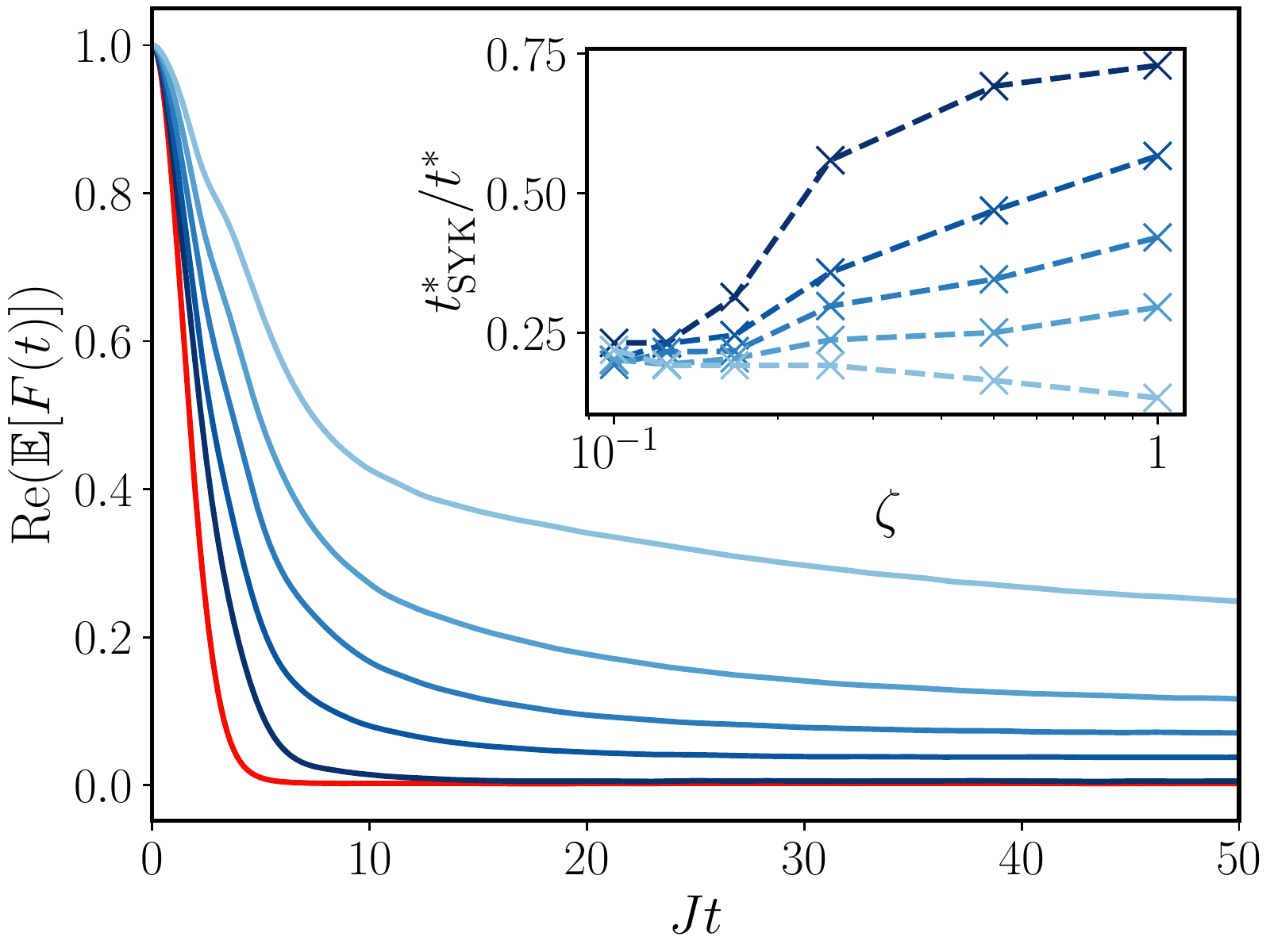}
    \caption{\textbf{OTOCs of effective and target model.}
        Dynamics of $\mathrm{Re}(\mathds{E}[F(t)] )$ for $\beta=0$ (see text) generated by $ H_{\mathrm{eff}}^{(4)} $ (blue curves) for $\zeta=1$, compared to those of $ H_\mathrm{SYK} $ with  $ J_{i_1 i_2 ; j_1 j_2}  \in \mathbb{R}$ (red curve).
        Data are for $N=10$ fermionic modes at half filling, and represent ensemble averages over $250$($1000$) disorder realizations of $ H_{\mathrm{eff}}^{(4)} $($ H_\mathrm{SYK} $).
        Shades of blue, from light to dark, correspond to $\tilde{\delta \omega}=10,2,1,1/2,1/10$.
        The inset shows the inverse times $1/t^*$, relative to that of the target model $1/t^*_{\mathrm{SYK}}$, at which $\mathrm{Re}(\mathds{E}[F(t^*)] )=1/e$, for each value of $\tilde{\delta \omega}$ (same shades of blue as main figure), 
        as a function of the transverse size $\zeta$ of the atomic cloud.
        With decreasing $\tilde{\delta \omega}$ and increasing $\zeta$, the OTOCs approach the fast dynamics of the SYK model.
    }
    \label{f:otocs}
\end{figure}
Figure~\ref{f:otocs} shows the real part of the  OTOC ${ F(t)=\tr(\rho_\beta W^\dagger(t)V^\dagger W(t) V) }$, calculated with respect to the infinite temperature state $\rho_{\beta=0} \propto \mathds{1}$, for a large atomic cloud $\zeta = 1$, and unitary operators $W=2 c_i^\dagger c_i-1$, $V=2 c_j^\dagger c_j-1$,  with $i=0$ and $j=1$.
The blue curves correspond to disorder averaged dynamics for $\tilde{\delta \omega} \in [0.1,10]$.
As $\tilde{\delta \omega}$ is decreased (effective number of involved cavity modes is increased) the OTOCs decay faster, and finally approach those of the SYK model (red).
This speed-up is more prominent for large transverse sizes $\zeta$ of the atomic cloud, as shown by the inset.
There, we use $1/t^*$---defined via $\mathrm{Re}(\mathds{E}[F(t^*)] ) = 1/e$, with $e$ being Euler's constant---as a proxy for the decay rate, and find the speed-up with decreasing $\tilde{\delta \omega}$ to be $\zeta$ dependent (fitting procedures such as that used in Refs.~\cite{Shen_etal2017, LantagneHurtubise_etal2020} proved to be unstable).
This dependence is due to the shape of the cavity mode functions:
As they feature a length scale (the cavity waist $w_0$), all interactions mediated by different transverse cavity modes become linearly dependent for small sizes, leading to a reduction of the effective number of modes mediating interactions in $ H_{\mathrm{eff}}^{(4)} $.
As a consequence, the convergence to the SYK model is slower.
This indicates that the effective model resembles the SYK model at a length scale comparable to the cavity waist, $\zeta =1$.

\begin{figure}[h]
    \centering
    \includegraphics[width=\linewidth]{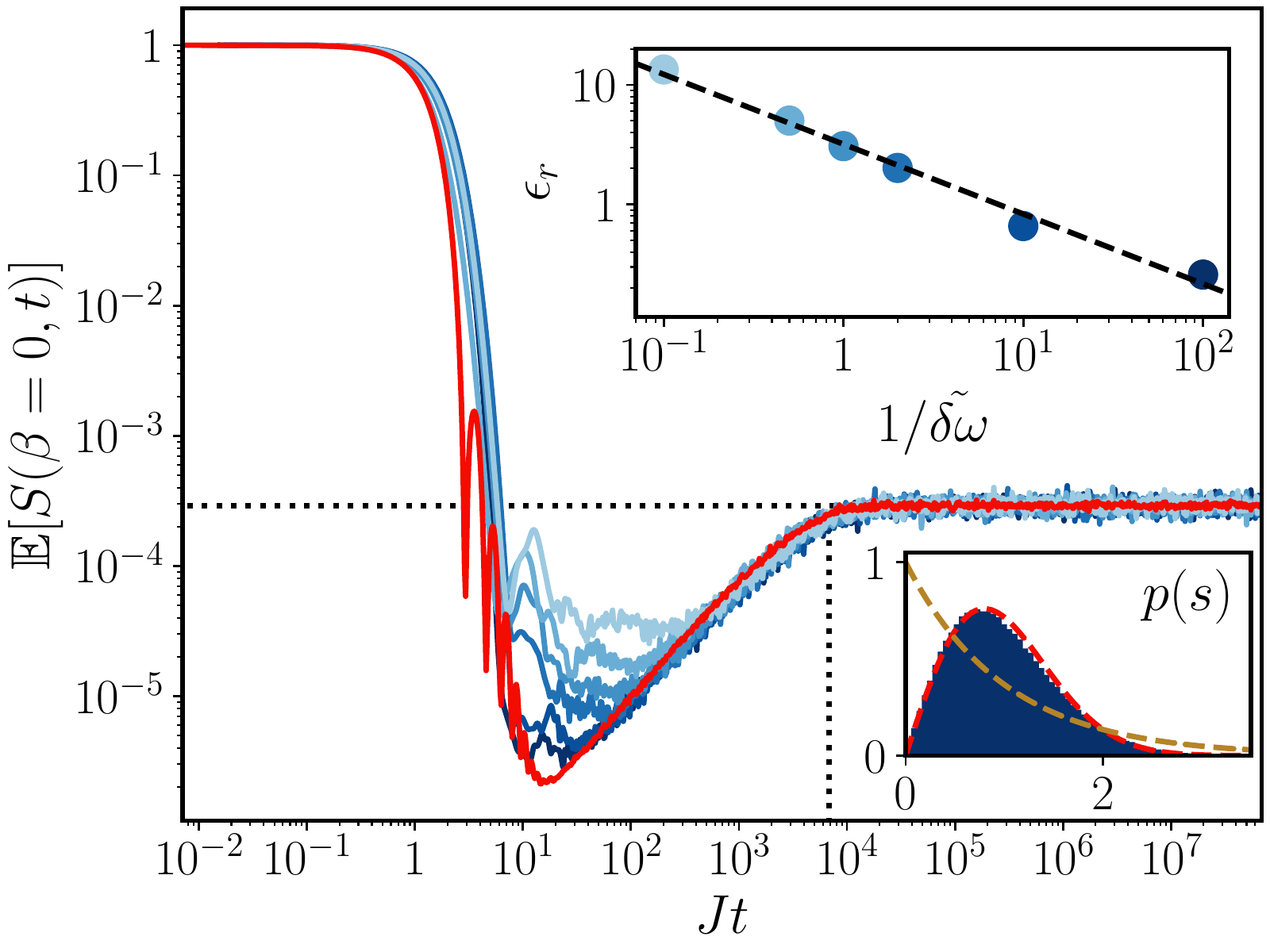}
    \caption{\textbf{SFF of effective and target model.}
        Disorder averaged SFF $\mathds{E}[S(\beta=0,t)]$ of $ H_{\mathrm{eff}}^{(4)} $ (blue curves) for $\zeta=1$, compared to that of $ H_\mathrm{SYK} $ with  $ J_{i_1 i_2 ; j_1 j_2}  \in \mathbb{R}$ (red line).
        Data are for $N=14$ fermionic modes at half filling, and represent ensemble averages over $100$($1000$) disorder realizations of $ H_{\mathrm{eff}}^{(4)} $($ H_\mathrm{SYK} $).        
        The Heisenberg time $t_\mathrm{H}=2D$(plateau height $1/D$) is indicated by the vertical(horizontal) black dotted line, showing very good agreement with the SFF of $ H_\mathrm{SYK} $.
        For the effective model, shades of blue from light to dark correspond to $\tilde{\delta \omega}=10,2,1,1/2,1/10,1/100$.
        The time axes for the SFFs of $ H_{\mathrm{eff}}^{(4)} $ have been rescaled in order to match the Heisenberg time of $ H_\mathrm{SYK} $ (see Methods, Sec.~\ref{s:numsim}).
        The upper inset shows that the deviation $\epsilon_r$ (see text) of the ramp time $t_\mathrm{r}$ decreases as a power law $\tilde{\delta \omega}^{-\alpha}$ with $\alpha=0.58 \pm 0.03$, as extracted from a least-squares fit (black dashed line).
        The lower inset shows the level-spacing distribution of $ H_{\mathrm{eff}}^{(4)} $ at $\tilde{\delta \omega}=0.01$ (same shade of blue as main figure), which agrees well with the Wigner surmise for the Gaussian orthogonal ensemble (red dashed line).
        Poisson level statistics (brown dashed line) are shown for comparison.
    }
    \label{f:sff}
\end{figure}
A typical diagnostic of quantum chaos is the distribution $p(s)$ of the spacings $s$ of nearest-neighbour energy-levels (Fig.~\ref{f:sff}, lower inset), with integrable models displaying Poisson statistics (brown dashed curve) and chaotic models following the Wigner surmise (red dashed curve, for the Gaussian Orthogonal Ensemble) \cite{Brody_etal1981}.
Indeed, the SYK model is known to follow the Wigner surmise \cite{You_etal2017}, and we find that also for $ H_{\mathrm{eff}}^{(4)} $, $p(s)$ matches the Wigner surmise very well for all considered values of $\tilde{\delta \omega}$ (histogram of inset. See also Fig.~\ref{f:convergence_test}e).
In contrast to the OTOCs, there is no qualitative change as a function of $\tilde{\delta \omega}$.
However, since $p(s)$ is a measure of short range (nearest-neighbour) spectral correlations, it is only able to identify the presence of chaotic behaviour at long timescales, on the order of $1/s$.
In order to study the many-body chaotic properties of our effective model in more detail, we thus turn to the SFF, which is sensitive also to long-range spectral correlations, and thus able to diagnose chaotic behaviour already at short to intermediate times.
Fig.~\ref{f:sff} shows the SFF $S(\beta,t) = \abs{ Z(\beta + it) / Z(\beta) }^2$ [where $Z(\beta + it)$ is the analytically continued partition function] of $ H_{\mathrm{eff}}^{(4)} $ for $\zeta=1$, as compared to that of $ H_\mathrm{SYK} $, at inverse temperature $\beta=0$.
The SFF of $ H_\mathrm{SYK} $ (red curve) is known to exhibit the characteristic features predicted by random matrix theory (RMT) \cite{Cotler_etal2017_1, delCampo_etal2017, Cotler_etal2017_2, SonnerVielma_2017, AltlandBagrets_2018, delCampo_etal2018, Altland_etal2021}: Early time oscillations decaying as $t^{-3}$, followed by a ramp linear in $t$, which transitions to a plateau of value $1/D$ at the Heisenberg time $2D$ ($D$ is the Hilbert space dimension).
For all values of $\tilde{\delta \omega}$ considered, the SFF of $ H_{\mathrm{eff}}^{(4)} $ (blue curves) shows signatures of the dip and ramp and, similar to the OTOCs, the depth of the dip and the temporal range of the ramp  approach that of the SYK model as $\tilde{\delta \omega}$  is decreased (light to dark blue).

For a quantitative comparison of the effective and target model, one would ideally want to compare their Thouless time.
This, however, is generically masked by the early-time slope, making it very challenging to determine numerically even for the target model \cite{Gharibyan_etal2018}.
Instead, we focus on comparing the time $t_\mathrm{r}$ at which the ramp of the SFF starts.
To extract this time, we follow the procedure of Ref.~\cite{Gharibyan_etal2018} by defining $t_\mathrm{r}$ as the earliest time at which the deviation of the SFF from a linear fit to the ramp falls below a chosen threshold (see Fig.~\ref{f:convergence_test}f~and~g for an example).
The upper inset to Fig.~\ref{f:sff} shows that the deviation $\epsilon_r = \abs{t_\mathrm{r}(\tilde{\delta \omega}) - t_\mathrm{r,SYK} } / t_\mathrm{r,SYK}$ scales as $\tilde{\delta \omega}^{-\alpha}$, with $\alpha \approx 0.58$.

These results highlight the importance of the microscopic parameters $\tilde{\delta \omega}$ and $\zeta$ for laboratory implementations using cQED. Using them, the interactions of Eq.~\eqref{e:Jmnpq_asym_main} can be tuned into a regime (small $\tilde{\delta \omega}$, large $\zeta$) in which the fast scrambling dynamics (OTOCs) and long-range spectral correlations (SFF) of the SYK model are well approximated. 
Finally, whilst here we have considered the ratio of speckle correlation length $\xi$ to oscillator length $ x_0 $ to be fixed (see Methods, Sec.~\ref{s:numsim}), tuning also $\xi/ x_0 $ may offer the capability to further optimise the chaotic and scrambling properties of $H_\mathrm{eff}$.

\subsection{Experimental implementation}
Our simulations highlight the key requirements for $H_\mathrm{eff}$ to approximate faithfully the dynamics of $ H_\mathrm{SYK} $.
We now show that these can be met using existing experimental capabilities. 

The need for a large $\zeta$ suggests the use of a light atomic species. 
 $^6$Li  is a natural choice of Fermionic atom: considering a trap with transverse frequencies of $\SI{25}{\Hz}$, such as that produced by the magnetic field curvatures in Ref.~\cite{Roux_etal2021}, yields $ x_0 = \SI{5.8}{\micro\meter}$.
Quantum degenerate Fermi gases of  $^6$Li  are now routinely produced in high finesse cavities \cite{Roux_etal2020,Zhang_etal2021}.
To produce mesoscopic samples of ten to hundreds of atoms, the methods of Ref.~\cite{Serwane_etal2011} could be adapted to the context of cavity QED.
While the target model is amenable to direct numerical studies for small $N$, and becomes analytically tractable when perturbing around the $N\rightarrow\infty$ limit, this mesoscopic regime remains inaccessible by these methods.
At the same time this regime harbours highly interesting quantum effects both from the many-body and the holographic perspective, sensitive to non-perturbative effects caused by the presence of a finite level spacing.

The lowest possible cavity waists together with low mode spacing $\delta \omega$ are achieved through the use of close-to-concentric cavities.
For instance, Ref.~\cite{Sauerwein_etal2022} operates with  $^6$Li  in a cavity with a waist of $\SI{13.2}{\micro\meter}$, and mode waists as low as $\SI{2.4}{\micro\meter}$ have been demonstrated close to the concentric limit \cite{Nguyen_etal2018}. 
The coupling of atoms to a very large number of modes has been achieved in confocal cavities \cite{Vaidya_etal2018}.
In close-to-concentric cavities, transverse modes are not degenerate even close to the stability limit, but as long as the transverse mode spacing is much lower than the free-spectral range, it is possible to emulate multimode driving using a comb of pump frequencies, each tuned close to one transverse mode family \cite{Johansen_etal2022}.
While scaling-up the number of cavity modes together with atom number is challenging in this context, the high degeneracy of high-order cavity modes alleviates the experimental overheads, so that reaching up to several hundreds of atoms seems realistic.

An intrinsic limitation of cavity-QED platforms is the occurrence of dissipation channels in the form of spontaneous photon scattering and photon leakage through the cavity mirrors.
The corresponding irreversible dynamics are described by jump operators in the Lindblad equation formulation. 
These jump operators inherit the random structure of the light-matter coupling, formally reproducing a dissipative SYK model similar to that studied recently in Ref.~\cite{Lucas_etal2022} (see SI, Sec.~\ref{s:dissipation}).
The finite cooperativity of the cavity will yield a timescale below which the dynamics will be faithfully described by the Hamiltonian of Eq.~\eqref{e:Heff_main}.
Since the central feature of the SYK model is the onset of chaos at logarithmically short time, we expect the study of this process not to be strongly hindered by dissipative effects.\\

\section{Discussion}
We have shown that state-of-the-art cQED experiments---with ultracold fermionic atoms coupled to a multi-mode optical cavity, and subjected to a spatially disordered AC-Stark shift---are able to realise an effective model $H_\mathrm{eff}$ with dominant random all-to-all two-body interactions $ \mathcal{J}_{i_1 i_2 ; j_1 j_2} $, as given by Eq.~\eqref{e:Jmnpq_asym_main}.
Numeric calculations from first principles reveal an underlying Cauchy distribution, rather than the Gaussian distribution of the ideal target model $ H_\mathrm{SYK} $ in Eq.~\eqref{e:Hsyk}, and we have shown this feature to be present also in previous proposals based on solid-state and optical-lattice architectures. 
Nevertheless, our exact simulations show good agreement between the considered OTOCs and the SFF of the effective model and the target $ H_\mathrm{SYK} $.
We identify how one can tune into the relevant parameter regime as the number of cavity-modes mediating $ \mathcal{J}_{i_1 i_2 ; j_1 j_2} $ is increased, by decreasing the ratio of the mode-spacing to the cavity--drive detuning $\tilde{\delta \omega}=\delta \omega /  \Delta_{\mathrm{cd}} $.

Our work provides a guideline for realizing the SYK model in current cQED experiments. Even more, it offers exciting prospects for further exploration.
An immediate extension is to vary the disordered light shift in time, thereby generating random time-dependent interactions $ \mathcal{J}_{i_1 i_2 ; j_1 j_2} (t)$ as required for so-called Brownian SYK models \cite{Suenderhof_etal2019}. 
The setup also permits for controlled deformations of the model, opening a platform on which to test the robustness of SYK-type physics, a question that is attracting increased interest lately \cite{GarciaGarcia_etal2018,Krajewski_etal2019,Cao_etal2020,GarciaGarcia_etal2021,Fremling_etal2022,Tezuka_etal2023}.
Finally, whilst we have considered a single atomic cloud, multiple quasi-two-dimensional clouds can be trapped within the same cavity, and could be subjected to identical disorder using spatial light modulators.
Such an extension of our work may offer the exciting prospect of realizing coupled SYK systems such as described by the Maldacena--Qi model \cite{MaldacenaQi_2018}, and studied in the context of traversable wormholes \cite{GaoJafferis_2021, Kobrin_etal2023, Jaferris_etal2022}.

%%%%%%%%%%%%%%%%%%%%%%%%%%%%%%%%%%%%%%%%%
\bibliography{refs_syk_cavity_proposal}
%%%%%%%%%%%%%%%%%%%%%%%%%%%%%%%%%%%%%%%%%

\section*{Acknowledgements}
We thank the authors of Refs.~\cite{Chen_etal2018}~and~\cite{WeiSedrakyan_2021} for providing the data utilised in Fig.~\ref{f:histograms}b--d.
We acknowledge helpful discussions with Alberto Biella, Tigrane Cantat-Moltrecht, Ricardo Costa~de~Almeida, Tobias Donner, Tilman Esslinger, Alicia~J.~Koll\'{a}r and Francesca Orsi.
P.U., S.B., and P.H. acknowledge funding from Provincia Autonoma di Trento, and by Q@TN, the joint lab between University of Trento, FBK-Fondazione Bruno Kessler, INFN-National Institute for Nuclear Physics, and CNR-National Research Council.
This project has received funding from the European Research Council (ERC) under the European Union’s Horizon $2020$ research and innovation programme (grant agreement No $804305$).
S.B. acknowledges CINECA for the use of HPC resources under ISCRA-C project ISSYK-2 (HP10CP8XXF). This work has been supported in part by the Fonds National Suisse de la Recherche Scientifique (Schweizerischer Nationalfonds zur F\"orderung der wissenschaftlichen Forschung) through Project Grant 200020182513, and the NCCR 51NF40-141869, The Mathematics of Physics (SwissMAP).
J.P.B and N.S. acknowledge funding from the Fonds National Suisse de la Recherche Scientifique (Schweizerischer Nationalfonds zur F\"orderung der wissenschaftlichen Forschung) through Project Grant No 200021184654 and the Swiss State Secretariat for Education, Research and Innovation through grant MB22.00063.

\section{Methods}

\setcounter{figure}{0} 
\renewcommand{\thefigure}{E\arabic{figure}} 
\renewcommand{\figurename}{EXTENDED DATA FIG.}

\subsection{Details for derivation of effective model}\label{s:derivation_Heff}
Here we summarise the steps performed to derive the effective model $H_\mathrm{eff}$ presented in the main text.
We begin by repeating the microscopic many-body Hamiltonian describing the cQED implementation of the main text (with $\hbar = 1$),
\begin{equation}\label{e:sm_Hmb}
    \begin{split}
        H_\mathrm{mb} =& \sum_m \omega_m a^\dagger_m a_m + \int d^2r \omega_\mathrm{a} (\boldsymbol{r}) \psi_\mathrm{e}^\dagger (\boldsymbol{r}) \psi_\mathrm{e} (\boldsymbol{r}) \\
        &+ \frac{1}{2} \sum_m \int d^2r \left( \Omega_m g_m(\boldsymbol{r})  a_m \psi_\mathrm{e}^\dagger(\boldsymbol{r}) \psi_\mathrm{g}(\boldsymbol{r}) + \mathrm{H.c.} \right) \\
        &+ \Omega_\mathrm{d} \int d^2r \left( g_\mathrm{d}(\boldsymbol{r}) e^{-i\omega_\mathrm{d} t} \psi_\mathrm{e}^\dagger(\boldsymbol{r}) \psi_\mathrm{g}(\boldsymbol{r})  + \mathrm{H.c.} \right) .
    \end{split}
\end{equation}
We note the following:
first, we have allowed for a disordered atomic resonance frequency $\omega_\mathrm{a} (\boldsymbol{r})$ (see SI, Sec.~\ref{s:acStark} for how this can be induced by a speckled light shift).
Second, we have dropped the contributions from the atoms' motional degrees of freedom $ H_{\mathrm{kt}} $, as it does not 
alter any of the following transformations, so we can add it back in at the end of the derivation of $H_\mathrm{eff}$ (for an explicit demonstration, see SI, Sec.~\ref{s:Hkt_in_derivation}).
Third, since we consider a two-dimensional atomic cloud trapped at an antinode of the longitudinal cavity mode,
the relevant spatial variation of the cavity modes is that of their transverse profiles.
These are given by Hermite--Gauss modes, which are labelled by transverse-mode indices $n_x, n_y \geq 0$, and have frequency $\omega_\mathrm{c} + (n_x+n_y) \delta \omega$, where $\omega_\mathrm{c}$ is the frequency of the lowest cavity mode, and $\delta \omega$ is the transverse-mode spacing.
We use Cantor's pairing function to assign a unique integer label $m=\mathrm{CP}(n_x,n_y) \geq 0$ to each transverse cavity mode.
Introducing the notation $m_\Sigma \equiv n_x+n_y$ ($n_x,n_y$ can be obtained from $m$ by inverting the pairing function), we write the frequencies of the transverse cavity modes (TCM) are $\omega_m = \omega_\mathrm{c} + m_\Sigma \delta \omega$.
Finally, we assume that the drive beam, of amplitude $g_\mathrm{d}(\boldsymbol{r})$, propagates transverse to the cavity($z$)-axis, and directly interacts with the fermionic atoms [last term of Eq.~\eqref{e:sm_Hmb}].
Depending on the transverse size $\zeta$ of the atomic cloud, the long wavelength approximation utilized in the main text may thus not be valid.
In such a case one may consider reducing the angle between the drive's wave-vector and the cavity axis, in order to increase the effective (projected) wavelength over the cloud.
This scenario is captured by the derivation of this section, as it amounts to modifying $g_\mathrm{d}(\boldsymbol{r})$.
However, this approach is limited by the cavity geometry, which may prohibit a sufficient enhancement of the effective wavelength.
As an alternative, one may consider an on-axis drive, for which the derivation proceeds analogously to that presented here, and is summarised in Sec.~\ref{s:lgtdl_drive} of the SI.

\textbf{Rotating frame.---}
Going into the rotating frame (RF) generated by
$H_\mathrm{RF} = \omega_\mathrm{d} \int d^2r \psi_\mathrm{e}^\dagger (\boldsymbol{r}) \psi_\mathrm{e} (\boldsymbol{r}) + \omega_\mathrm{d}  \sum_m  a^\dagger_m a_m$,
we obtain the time-independent Hamiltonian
\begin{equation}\label{e:H_tindep}
    \begin{split}
        H =&  \sum_m   \Delta_m  a^\dagger_m a_m -  \int d^2r  \Delta_{\mathrm{da}} (\boldsymbol{r}) \psi_\mathrm{e}^\dagger (\boldsymbol{r}) \psi_\mathrm{e} (\boldsymbol{r})\\
        &+ \int d^2r \left( \Phi(\boldsymbol{r}) \psi_\mathrm{e}^\dagger(\boldsymbol{r}) \psi_\mathrm{g}(\boldsymbol{r})  + 
        \Phi^\dagger(\boldsymbol{r}) \psi_\mathrm{g}^\dagger(\boldsymbol{r}) \psi_\mathrm{e}(\boldsymbol{r}) \right).
    \end{split}
\end{equation}
Here, $ \Delta_{\mathrm{da}} (\boldsymbol{r}) \equiv \omega_\mathrm{d} - \omega_\mathrm{a}(\boldsymbol{r})$($ \Delta_m  \equiv  \Delta_{\mathrm{cd}}  + m_\Sigma \delta \omega$) is the drive--atom(cavity--drive) detuning, where $ \Delta_{\mathrm{cd}}  \equiv \omega_\mathrm{c} - \omega_\mathrm{d}$, and
\begin{equation}\label{e:Phi}
    \Phi(\boldsymbol{r}) = \Omega_\mathrm{d} g_\mathrm{d}(\boldsymbol{r}) + \frac{1}{2} \sum_m  \Omega_m g_m(\boldsymbol{r}) a_m .
\end{equation}

\textbf{Adiabatic elimination of the atomic excited states.---}
Assuming $\abs{  \Delta_{\mathrm{da}} (\boldsymbol{r}) }$ to be the dominant energy scale at all $\boldsymbol{r}$, the system is in the dispersive regime (low saturation limit) \cite{MetcalfcanDerStraten2002}, and  $\psi_\mathrm{e}(\boldsymbol{r})$ adiabatically follows $\psi_\mathrm{g}(\boldsymbol{r})$ according to
\begin{equation}\label{e:psi_e_adelim_methods}
    \psi_\mathrm{e}(\boldsymbol{r}) = \frac{\Phi(\boldsymbol{r}) 
        \psi_\mathrm{g}(\boldsymbol{r})}{ \Delta_{\mathrm{da}} (\boldsymbol{r}) } .
\end{equation}
Inserting Eq.~\eqref{e:psi_e_adelim_methods} into the Heisenberg equations of motion for $a_m$ and $\psi_\mathrm{g}(\boldsymbol{r})$, one can determine the corresponding Hamiltonian to be
\begin{equation}\label{e:H_adelim}
    H =  \sum_m    \Delta_m  a^\dagger_m a_m + \int d^2r \frac{ \Phi^\dagger(\boldsymbol{r}) \Phi(\boldsymbol{r}) \psi_\mathrm{g}^\dagger(\boldsymbol{r}) \psi_\mathrm{g}(\boldsymbol{r}) }{  \Delta_{\mathrm{da}} (\boldsymbol{r}) } .
\end{equation}
In what follows, we simplify our notation by denoting the remaining field operator $\psi_\mathrm{g}(\boldsymbol{r})$ as $\psi(\boldsymbol{r})$.

\textbf{Schrieffer--Wolff transformation.---}
We group the Hamiltonian of Eq.~\eqref{e:H_adelim} as $H = H_0 + V$, where
\begin{align}
    H_0 =&  \sum_m    \Delta_m  a^\dagger_m a_m + \int d^2r \frac{ \abs{\Omega_\mathrm{d} g_\mathrm{d}(\boldsymbol{r})}^2  }{ \Delta_{\mathrm{da}} (\boldsymbol{r})} \psi^\dagger(\boldsymbol{r}) \psi(\boldsymbol{r}) , \label{e:SWT_H0} \\
    V =&  \frac{1}{4} \int d^2r \sum_{m, n} \left( \Omega_m^* g_m^*(\boldsymbol{r}) \Omega_n g_n (\boldsymbol{r})a^\dagger_m a_n \right)  \frac{\psi^\dagger(\boldsymbol{r}) \psi(\boldsymbol{r})}{ \Delta_{\mathrm{da}} (\boldsymbol{r})}  \nonumber  \\
    &+  \sum_m  
    ( a_m \Theta_m + \mathrm{H.c.})  , \text{ with}  \label{e:SWT_V} \\
    \Theta_m =& \frac{1}{2} \int d^2r \frac{ \Omega_\mathrm{d}^* g_\mathrm{d}^*(\boldsymbol{r}) \Omega_m g_m(\boldsymbol{r}) \psi^\dagger(\boldsymbol{r}) \psi(\boldsymbol{r}) }{ \Delta_{\mathrm{da}} (\boldsymbol{r})}  .
\end{align}
The effective two-body interactions of Eq.~\eqref{e:Jmnpq_asym_main} are obtained by decoupling the atom--light interactions contained in $V$.
We do so through a Schrieffer--Wolff transformation $ e^S H e^{-S}=H_0+ \left( V + \comm{S}{H_0} \right) + \comm{S}{V} + \frac{1}{2} \comm{S}{\comm{S}{H_0 + V}} + \ldots $, with the anti-Hermitian generator $S$ chosen such that $V + \comm{S}{H_0}= 0$.

The transformation is simplified by using that $ \abs{\Omega_m} \ll \abs{\Omega_\mathrm{d}}, \, \forall m$ (see SI, Sec.~\ref{s:energies}).
Within this hierarchy of scales, $V$ can be approximated as $V \simeq  \sum_m  (a_m \Theta_m + \mathrm{H.c.} )\,$ [we will return to this simplification below Eq.~\eqref{e:H_post_SWT}].
The decoupling is then achieved by choosing the generator
\begin{equation}\label{e:SWT_generator_methods}
    S = -   \sum_m  \left( \frac{ 1 }{   \Delta_m } a_m \Theta_m - \mathrm{H.c.} \right) ,
\end{equation}
which is of order $\abs{\Omega_\mathrm{d} \Omega_m} / \abs{ \Delta_{\mathrm{da}}   \Delta_{\mathrm{cd}} }$.
Truncating $H_\mathrm{eff}$ at $\mathcal{O}(S)$  (which requires $\abs{\Omega_\mathrm{d} \Omega_m} / \abs{ \Delta_{\mathrm{da}}   \Delta_{\mathrm{cd}} } \ll 1$), then yields
\begin{equation}\label{e:H_post_SWT}
    H_\mathrm{eff} = H_0 + \frac{1}{2}\comm{S}{V} = H_0 -  \sum_m  \frac{\Theta_m^\dagger \Theta_m}{  \Delta_m } .
\end{equation}
The last term is the effective two-body interaction, mediated by the exchange of virtual photons between pairs of atoms located at arbitrary positions, see Fig.~\ref{f:concept}c. 
The remaining photonic contribution within $H_0$ is eliminated by projecting onto a subspace with a fixed number of cavity photons.

We briefly return to the first term of $V$, which we denote as $V'$:
If the assumption $ \abs{\Omega_m} \ll \abs{\Omega_\mathrm{d}}, \, \forall m$ is not met, then [using the same generator $S$ of Eq.~\eqref{e:SWT_generator_methods}, and again truncating at $\mathcal{O}(S)$] one would obtain the additional terms $V' + \comm{S}{V'}$ in Eq.~\eqref{e:H_post_SWT}.
Projection onto a subspace of fixed photon number would then remove the commutator (as it is linear in photonic operators), and the remaining term would simply modify $H_0$.
In what follows, we continue to neglect this term.

The one-body part of the Hamiltonian in Eq.~\eqref{e:H_post_SWT} is given by
\begin{equation}\label{e:H_1body}
     H_{\mathrm{kt}}  + \int d^2r \frac{  \abs{\Omega_\mathrm{d} g_\mathrm{d}(\boldsymbol{r})}^2  }{ \Delta_{\mathrm{da}} (\boldsymbol{r})} \psi^\dagger(\boldsymbol{r}) \psi(\boldsymbol{r}) ,
\end{equation}
where we have added the kinetic and external trap terms back in, which just pass through the various transformations done from Eq.~\eqref{e:sm_Hmb} up to this point, see SI, Sec.~\ref{s:Hkt_in_derivation}.
This one-body Hamiltonian can be diagonalised to the form $\sum_{i_1}\epsilon_{i_1} c^\dagger_{i_1} c_{i_1}$. 
Expanding the field operators in Eq.~\eqref{e:H_post_SWT} in the corresponding eigenbasis as
$\psi(\boldsymbol{r}) = \sum_{i_1} \phi_{i_1}(\boldsymbol{r}) c_{i_1}$
then yields
\begin{equation}\label{e:Heff}
    \begin{split}
        &H_{\mathrm{eff}} = \sum_{i_1}\epsilon_{i_1} c^\dagger_{i_1} c_{i_1} \\
        &+  \sum_m  \frac{1}{ \Delta_m } \left( \sum_{i_1,j_1}  I_{i_1 j_1, m}  c^\dagger_{i_1} c_{j_1} \right) \!\! \left( \sum_{i_2,j_2} I_{j_2 i_2, m} c^\dagger_{j_2} c_{i_2} \right)^\dagger ,
    \end{split}
\end{equation}
where 
\begin{equation}\label{e:interaction_integral}
    I_{i_1 j_1, m} = \frac{1}{2} \int d^2r \frac{ \Omega_\mathrm{d}  g_\mathrm{d}(\boldsymbol{r}) \left( \Omega_m g_m(\boldsymbol{r}) \right)^* \phi^*_{i_1}(\boldsymbol{r}) \phi_{j_1}(\boldsymbol{r}) }{ \Delta_{\mathrm{da}} (\boldsymbol{r})} .
\end{equation}
The two-body term of the effective Hamiltonian given in Eq.~\eqref{e:Heff}---written as a sum over products of fermion bilinear operators---is of the form of a low (matrix) rank variant of the SYK model \cite{Kim_etal2020} in which the desired rank-$4$ interaction tensor decomposes as a sum over products of rank-$2$ tensors,
\begin{equation}\label{e:Jmnpq}
    \mathcal{J}_{i_1 i_2 ; j_1 j_2} = \sum_{m} \frac{ I_{i_1 j_1, m} I_{j_2 i_2, m}^* }{ \Delta_m } = \mathcal{J}_{j_2 j_1 ; i_2 i_1}^* .
\end{equation}
It was shown in Ref.~\cite{Kim_etal2020} that for extensive matrix rank the above interaction tensor can realise a maximally scrambling model, nearly indistinguishable from the target SYK model of Eq.~\eqref{e:Hsyk}.
In our proposal, the rank is determined by the number of TCMs mediating the interaction, and this can in turn be controlled by tuning the ratio $\tilde{\delta \omega} = \delta \omega/ \Delta_{\mathrm{cd}} $ of the transverse-mode spacing $\delta \omega$, to the cavity--drive detuning $ \Delta_{\mathrm{cd}} $.
This is most evident when extracting the overall energy scale (denoting the detuning of the drive from the bare atomic resonance as $ \Delta_{\mathrm{da}}  \equiv \omega_\mathrm{d} - \omega_\mathrm{a}$)
\begin{equation}\label{e:energy_scale_J}
    \mathcal{E} = \frac{1}{ \Delta_{\mathrm{cd}} } \frac{\Omega_\mathrm{d}^2}{ \Delta_{\mathrm{da}} } \frac{\Omega_{m=0}^2}{ \Delta_{\mathrm{da}} } 
\end{equation}
from the interactions of Eq.~\eqref{e:Jmnpq}, as
\begin{equation}\label{e:J2_rescaled}
    \mathcal{J}_{i_1 i_2 ; j_1 j_2} 
    = \mathcal{E} \tilde{\mathcal{J}}_{i_1 i_2 ; j_1 j_2}
    = \sum_{m} \frac{ \tilde{I}_{i_1 j_1, m} \left( \tilde{I}_{ j_2 i_2, m } \right)^* }{ 1 + m_\Sigma \frac{\delta \omega}{ \Delta_{\mathrm{cd}} } } ,
\end{equation}
where 
\begin{equation}\label{e:I_rescaled}
    \tilde{I}_{i_1 j_1, m} 
    = \frac{1}{2} \int d^2r \frac{1}{ \Delta_{\mathrm{da}} (\boldsymbol{r}) /  \Delta_{\mathrm{da}}  } g_\mathrm{d}(\boldsymbol{r}) g_{m}(\boldsymbol{r})^* \phi^*_{i_1}(\boldsymbol{r}) \phi_{j_1}(\boldsymbol{r}) .
\end{equation}
In $\mathcal{E}$, we have set $\Omega_m = \Omega_{m=0}$, since $m$-dependent corrections to $\Omega_m$ are of order $\delta \omega / \omega_\mathrm{c}$, and can thus be assumed to be negligible: for the setup of Ref.~\cite{Sauerwein_etal2022}, for instance, the corrections are of order $\delta \omega / \omega_\mathrm{c} \sim 10^{-6}$ since there, $\delta \omega \sim \SI{100}{\MHz}$ whilst $\omega_\mathrm{c} \sim \SI{100}{\THz}$.

In the main text, we study the normal-ordered form of $H_\mathrm{eff}$.
This yields an additional one-body term, of the same order as $ \mathcal{J}_{i_1 i_2 ; j_1 j_2} $, which is however suppressed as $1/N^2$ due to combinatorics, and we thus neglect this term.
The anti-symmetric contribution to the normal-ordered two-body interaction, summarised in Eq.~\eqref{e:Jmnpq_asym_main} of the main text, is thus given by
\begin{equation}\label{e:Jmnpq_asym}
    \begin{split}
        &\mathcal{J}_{i_1 i_2 ; j_1 j_2} = \frac{1}{2} \sum_{m} \frac{ \abs{ \Omega_\mathrm{d} \Omega_m }^2}{  \Delta_m } \\
        &\times \int d^2r \int d^2r' \frac{ \mathrm{Re} \left[ g_\mathrm{d}(\boldsymbol{r}) g_m(\boldsymbol{r})^* g_\mathrm{d}(\boldsymbol{r'})^* g_m(\boldsymbol{r'})  \right] }{  \Delta_{\mathrm{da}} (\boldsymbol{r})  \Delta_{\mathrm{da}} (\boldsymbol{r'}) } \\
        &\times \left( \phi_{i_1}(\boldsymbol{r})     \phi_{i_2}(\boldsymbol{r'})  -  \phi_{i_2}(\boldsymbol{r}) \phi_{i_1}(\boldsymbol{r'})  \right)^* \phi_{j_1}(\boldsymbol{r}) \phi_{j_2}(\boldsymbol{r'}) ,
    \end{split}
\end{equation}
which is real if $\phi_{k}(\boldsymbol{r}) \in \mathbb{R}$ for $k=i_1, i_2, j_1, j_2$.

\subsection{Numeric implementation}\label{s:numsim}
\begin{figure*}[h!]
    \centering
    \includegraphics{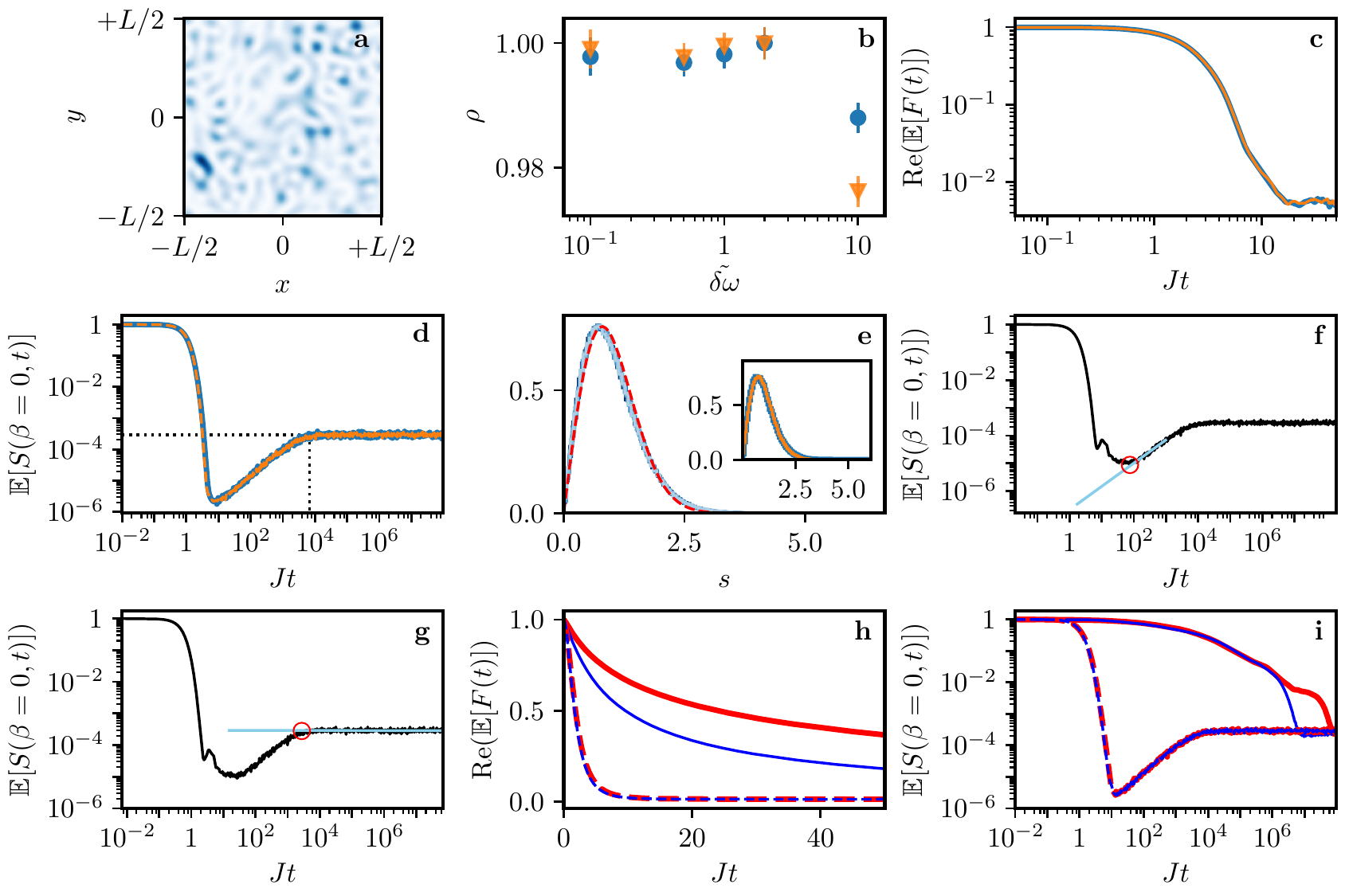} 
    \caption{\textbf{Details of numeric implementation.}
        (a) A realization of the speckled intensity distribution used to numerically define the disordered drive--atom detuning $ \Delta_{\mathrm{da}} (\boldsymbol{r})$.
        (b)--(e) Comparison of data for mode-cut-off $M=240$ (blue) versus $M=500$ (orange):
        (b) The interpolation parameter $\rho$ [see Eq.~\eqref{e:psv}], quantifying the shape of the probability distribution of $ \tilde{\mathcal{J}}_{i_1 i_2 ; j_1 j_2} $ for $\tilde{\delta \omega}=1/10,1/2,1,2,10$.
        Markers and error-bars, respectively, represent the mean and standard deviation of $\rho$ for an ensemble of $100$ disorder realizations.
        (c), (d) OTOC (operator choice as in main text), respectively, SFF for $N=10$($N=14$) fermionic modes at half-filling, with $\zeta=1$, and $\tilde{\delta \omega}=1/10$($\tilde{\delta \omega}=1/1000$), averaged over $250$($100$) disorder realizations.
        (e, main) Level-spacing distribution $p(s)$ of $ H_{\mathrm{eff}}^{(4)} $, for $N=14$ fermionic modes at half-filling and $\zeta=1$.
        For all considered $\tilde{\delta \omega}=10,2,1,1/2,1/10,1/100$ (light to dark blue), $p(s)$ agrees well with the Wigner surmise for the Gaussian Orthogonal Ensemble (red dashed curve).
        (e, inset) Convergence of $p(s)$, with respect to mode-cut-off $M$, of $ H_{\mathrm{eff}}^{(4)} $ for $N=14$ fermionic modes at half-filling, $\zeta=1$, and $\tilde{\delta \omega}=1/1000$.
        (f), (g) Examples of the relative error technique of Ref.~\cite{Gharibyan_etal2018}, used here to extract (f) the ramp time $t_\mathrm{r}$, and (g) the Heisenberg time $t_\mathrm{H}$ of the SFF (black curve) of Fig.~\ref{f:sff} for $\tilde{\delta \omega}=1.0$.
        A linear fit (sky-blue line) to the ramp, respectively plateau, is used to determine the earliest time at which the relative deviation $\varepsilon \equiv \abs{S(t)-f(t)} / \abs{f(t)}$ of the SFF $S(t)$ from the linear fit $f(t)$ falls below an error threshold.
        This time is designated as the ramp, respectively, Heisenberg time.
        Here, and for the data of the main text, we set the threshold to $0.01$, and the red circle markers are centred on $S(t=t_\mathrm{r})$, respectively, $S(t=t_\mathrm{H})$.
        For illustrative purposes, the linear fits have been plotted beyond their domain of validity, which is chosen by inspection for each data set.
        (h), (i)
        Simulations illustrating the qualitative differences for drive amplitudes with non-uniform phases.
        The OTOCs (h) and SFFs (i) are generated by $ H_{\mathrm{eff}}^{(4)} $ with interactions $ \tilde{\mathcal{J}}_{i_1 i_2 ; j_1 j_2} $ in which the drive amplitude has an oscillating phase $g_\mathrm{d}(\boldsymbol{r}) = \exp(i k_\mathrm{d} x)$, where $k_\mathrm{d}=2\pi/\lambda_\mathrm{d}$ is the drive beam's wave-number, and the drive propagates in the (transverse to the cavity axis) $x$-direction.
        OTOC(SFF) data are for a system of $N=10$($N=14$) fermionic modes at half-filling, averaged over $500$($200$) disorder realizations.
        Both panels  compare dynamics for different transverse atomic cloud sizes $\zeta=1$($\zeta=0.1$), indicated by solid(dashed) curves, and different $\tilde{\delta \omega}=100$($\tilde{\delta \omega}=1/1000$) shown in red(blue).
        Contrary to the case of homogeneous drive (main text), the dynamics in the smaller cloud ($\zeta=0.1$) are faster than in the larger cloud ($\zeta=1$).
        However, the data for different $\tilde{\delta \omega}$ collapse on top of one-another, so that no speed up is apparent as $\tilde{\delta \omega}$ is tuned.
        This indicates that the homogeneous drive utilised in the main text is an important ingredient for the proposal.
    }
    \label{f:convergence_test}
\end{figure*}
Here we provide details on the numeric simulation of $H_\mathrm{eff}$, as implemented for the data presented in this work.
In short, for a given realization of the speckle, we solve for the eigenmodes $\phi_{i}(\boldsymbol{r})$ of $H_0$, use these to approximate the interaction integrals of $\tilde{I}_{i_1 j_1, m}$ via Riemann sums.
From these, we construct the antisymmetric interactions $ \tilde{\mathcal{J}}_{i_1 i_2 ; j_1 j_2}  =  \mathcal{J}_{i_1 i_2 ; j_1 j_2}  / \mathcal{E}$, and finally diagonalise the normal-ordered two body part of $H_\mathrm{eff}$, $ H_{\mathrm{eff}}^{(4)}  / \mathcal{E} = \sum_{i_1, i_2, j_1, j_2}  \tilde{\mathcal{J}}_{i_1 i_2 ; j_1 j_2}  c^\dagger_{i_1} c^\dagger_{i_2} c_{j_1} c_{j_2}$.
The thus obtained spectrum is then used to simulate the dynamics of $ H_{\mathrm{eff}}^{(4)} $.

As motivated in Sec.~\ref{s:compensation} of the SI, we drop the dipole term of $H_0$ in Eq.~\eqref{e:H_1body}, such that 
\begin{equation}
    H_0= H_{\mathrm{kt}} = \int d^2r \psi_\mathrm{g}^\dagger(\boldsymbol{r}) \left( \boldsymbol{p}^2 / (2  m_{\mathrm{at}}  ) +  V_{\mathrm{t}} (\boldsymbol{r}) \right) \psi_\mathrm{g}(\boldsymbol{r}) ,
\end{equation}
which is simply a quantum harmonic oscillator (QHO) Hamiltonian for $ V_{\mathrm{t}} (\boldsymbol{r}) = ( m_{\mathrm{at}}  \omega_\mathrm{t}^2 /2) \boldsymbol{r}^2$ (we assume the trap to be isotropic in the plane transverse to the cavity axis).
Introducing dimensionless coordinates $\boldsymbol{r}' \equiv \boldsymbol{r}/ x_0 $, where $ x_0  = \sqrt{1 / ( m_{\mathrm{at}}  \omega_\mathrm{t})}$ is the zero-point fluctuation of the ground state of $ H_{\mathrm{kt}} $, we obtain
\begin{equation}\label{e:qho}
    H_0= \frac{\omega_\mathrm{t}}{2} \int d^2r' \psi_\mathrm{g}^\dagger(\boldsymbol{r}') \left( -(\vec{\nabla}')^2 + (\boldsymbol{r}')^2 \right) \psi_\mathrm{g}(\boldsymbol{r}') .
\end{equation}
The eigenmodes $\phi_i(\boldsymbol{r}')$ of the above QHO Hamiltonian are products of Hermite--Gauss modes ${ \phi_i(\boldsymbol{r}')=\psi_{n_x^{(i)} }(x') \psi_{n_y^{(i)}} (y') }$.
Nevertheless, we obtain the eigenmodes via exact diagonalization, to maintain flexibility of our numeric calculations.

We construct the matrix representation of the Hamiltonian of Eq.~\eqref{e:qho} in the position basis over a square grid of $N_x \times N_x$ coordinates $\boldsymbol{r}'$, centred at $\boldsymbol{r}'=(0,0)$, and set $\omega_\mathrm{t}=1$.
For the remainder of this section, we drop the prime notation, and all length scales are to be understood as expressed in units of $ x_0 $.
Using exact diagonalization, we obtain the $N$ energetically lowest eigenmodes $\lbrace \phi_i(\boldsymbol{r}) \rbrace_{i=0}^{N-1}$, where $N$ is the desired system size.
To prevent distortions of the eigenmodes, the grid diameter $L$ must be chosen sufficiently large:
The spatial variance of a given mode $\phi_i(\boldsymbol{r})=\psi_{n_x^{(i)} }(x) \psi_{n_y^{(i)}} (y)$ is (in units of $ x_0 $) given by $(n^{(i)}+1)$, where $n^{(i)} = \mathrm{max}(n_x^{(i)},n_y^{(i)})$.
In the main text, we consider systems of $N\leq14$, for which the largest variance is $(4+1)$, and we thus set $L =10\approx 5\times \sqrt{4+1}$.
We set the number of grid-points as $N_x=200$, for which the relative error in the energy of the highest mode is less than $0.1\%$.

Next, we use the set $\lbrace \phi_i(\boldsymbol{r}) \rbrace_{i=0}^{N-1}$ to calculate the $[N(N-1)/2]^2$ antisymmetrised, two-body interaction amplitudes $ \tilde{\mathcal{J}}_{i_1 i_2 ; j_1 j_2} $ as defined by Eqs.~\eqref{e:J2_rescaled}~and~\eqref{e:Jmnpq_asym}, for a given input value of $\tilde{\delta \omega} = \delta \omega /  \Delta_{\mathrm{cd}} $.  
The spatial integrals are approximated as Riemann sums over the above coordinate grid.
As motivated in the main text, we work in a long-wavelength approximation such that $g_\mathrm{d}(\boldsymbol{r})=1$.
Assuming the pancake to be placed at the centre of the cavity-axis ($z=0$), the cavity modes $g_m(\boldsymbol{r})$ are Hermite--Gauss modes $g_m(\boldsymbol{r})  = \psi_{n_x}(x) \psi_{n_y}(y)$, where $m$ is obtained from integers $n_x,n_y \geq0$ via Cantor's pairing function, and
\begin{equation}
    \psi_{n_x}\!(x) = \sqrt{ \!\! \frac{ \sqrt{2}/w_0 }{ \sqrt{\pi} 2^{n_x} n_x! } } \exp(\frac{ - x^2 }{ 2( w_0/\sqrt{2} )^2 } ) \! H_{n_x} \! \left( \frac{x}{ w_0/\sqrt{2}} \right) .
\end{equation}
The parameter $w_0$ is the cavity waist at centre, and ${ \zeta =  x_0  / (w_0 / \sqrt{2}) }$ quantifies the spatial extent of the fermionic modes $\phi_{i}(\boldsymbol{r})$ relative to this waist.
In Figs.~\ref{f:histograms}--\ref{f:sff}, we present data for a range of sizes $\zeta \in [0.1,1]$, which is implemented in our numerics by keeping $ x_0 =1$ fixed, and varying the cavity waist as $w_0 = \sqrt{2} \zeta$.
We set a mode-cut-off $m=0,1,\ldots,M$ in Eq.~\eqref{e:Jmnpq_asym} as $M=240$ (for a test of convergence with respect to $M$, see Fig.~\ref{f:convergence_test}b--e), and
ensure that $N_x$ is large enough such that the frequencies of all modes entering Eq.~\eqref{e:Jmnpq_asym} are sampled above the Nyquist rate.

Finally, to produce the disordered detuning $ \Delta_{\mathrm{da}} (\boldsymbol{r})$ of Eq.~\eqref{e:lightshift}, we numerically generate speckle patterns according to the method of Ref.~\cite{DuncanKirkpatrick_2008}.
We assume the blue-detuned scenario $\Delta_\mathrm{b} > 0$, and set the spatial average of $ \frac{ \abs{ \Omega_{\mathrm{b}}(\boldsymbol{r}) }^2 / (4 \abs{\Delta_\mathrm{b} } ) }{ \abs{  \Delta_{\mathrm{da}}  } } $ to unity.
The mean number of speckle grains per linear dimension of the grid is a tunable parameter in the numerics.
Physically, this number is determined by the speckle correlation length $\xi$.
As an example, the light-shifting beam and numerical aperture of the setup of Ref.~\cite{Sauerwein_etal2022} would yield $(w_0/\sqrt{2}) / \xi \approx 17$.
We thus set the average number of speckle grains per linear dimension of the pancake to $17$.
A realization is shown in Fig.~\ref{f:convergence_test}a.    

Having obtained the set of amplitudes $ \tilde{\mathcal{J}}_{i_1 i_2 ; j_1 j_2} $, we use them to construct the Fock-space representation of $ H_{\mathrm{eff}}^{(4)}  / \mathcal{E} = \sum_{i_1, i_2, j_1, j_2}  \tilde{\mathcal{J}}_{i_1 i_2 ; j_1 j_2}  c^\dagger_{i_1} c^\dagger_{i_2} c_{j_1} c_{j_2}$ within the half-filling sector.
The dynamics are then solved via exact diagonalization \cite{Bandyopadhyay_etal2021, Paviglianiti_etal2023}.

The above procedure is repeated multiple times, with independent speckle realizations, to obtain the ensemble-averaged data ($\mathds{E}[\cdot]$) presented in the main text.

From our simulations of the spectral form factor (SFF), we extract the ramp, respectively, Heisenberg time via the procedure used in Ref.~\cite{Gharibyan_etal2018}.
In short, we fit a linear function to the ramp, respectively, plateau of a given ensemble-averaged SFF, and then determine the earliest time at which the relative deviation of the SFF from this fit is below $1\%$.
An example is shown in Fig.~\ref{f:convergence_test}f~and~g.

All numerics presented in this work are done under the long-wavelength approximation for the amplitude of the transverse drive beam $g_\mathrm{d}(\boldsymbol{r})=1$ which, as discussed in the main text, is motivated by the assumption of being within the Lamb--Dicke regime $ k_\mathrm{d}  x_0   \ll 1$, 
where $k_\mathrm{d}=2\pi/\lambda_\mathrm{d}$, and $\lambda_\mathrm{d}$ is the wavelength of the transverse drive.
The case of an oscillating drive amplitude $g_\mathrm{d}(\boldsymbol{r})=\exp(i k_\mathrm{d} x )$, propagating along the (transverse) $x$ direction, is shown in Fig.~\ref{f:convergence_test}h~and~i, which shows realizations of the out-of-time-order correlators (OTOCs) considered in the main text, and the SFF, for small ($\zeta=0.1$) and large ($\zeta=1.0$) transverse sizes of the atomic cloud and $\tilde{\delta \omega}=1/1000,100$, with the latter interpolating between many and few modes contributing to $ \tilde{\mathcal{J}}_{i_1 i_2 ; j_1 j_2} $.

We end this section with the expression for the (pseudo)Voigt profile \cite{Wertheim_etal1974} which we use to fit a given realization of the set of $[N(N-1)/2]^2$ antisymmetrised interactions $\lbrace  \tilde{\mathcal{J}}_{i_1 i_2 ; j_1 j_2}  | i_1 > i_2, j_1 > j_2 \rbrace$ (see Fig.~\ref{f:histograms}, and Fig.~\ref{f:convergence_test}b).
It is given by a superposition of a Cauchy and a Gaussian probability density, both centred at $\bar{x}$, and sharing the same full-width-at-half-max $2\sqrt{2\ln(2) }\sigma$,
\begin{equation}\label{e:psv}
    \begin{split}
            f(x) =& \rho \left[  \frac{2\sqrt{2\ln(2)} \sigma }{2\pi}  \frac{1}{(x-\bar{x})^2 + ( 2\sqrt{2\ln(2)} \sigma /2)^2} \right] \\
            &+ (1-\rho) \left[ \frac{1}{\sqrt{2 \pi \sigma^2}} \exp(-\frac{(x-\bar{x})^2}{2\sigma^2}) \right] .
    \end{split}
\end{equation}
The interpolation parameter $\rho \in [0,1]$ can thus be used to quantify the extent to which a given realization $\lbrace  \tilde{\mathcal{J}}_{i_1 i_2 ; j_1 j_2}  | i_1 > i_2, j_1 > j_2 \rbrace$ deviates from the targeted Gaussian ($\rho=0$), towards a Cauchy ($\rho=1$) distribution.

%%%%%%%%%%%%%%%%%%%%%%%%%%%%%%%
% Supplementary Information
%%%%%%%%%%%%%%%%%%%%%%%%%%%%%%%
\clearpage
\onecolumngrid

\setcounter{page}{1}
\setcounter{equation}{0}
\setcounter{figure}{0}
\setcounter{section}{0}
\renewcommand{\thepage}{S\arabic{page}}
\renewcommand{\theequation}{S\arabic{equation}}
\renewcommand{\thefigure}{S\arabic{figure}}
\renewcommand{\thesection}{S\Roman{section}}
\renewcommand{\figurename}{SUPPLEMENTARY FIG.} 

\section{Supplementary Information: A cavity quantum electrodynamics implementation of the Sachdev--Ye--Kitaev model}

\subsection{Energy scales}\label{s:energies}
\begin{figure}[h!]
    \centering
    \includegraphics{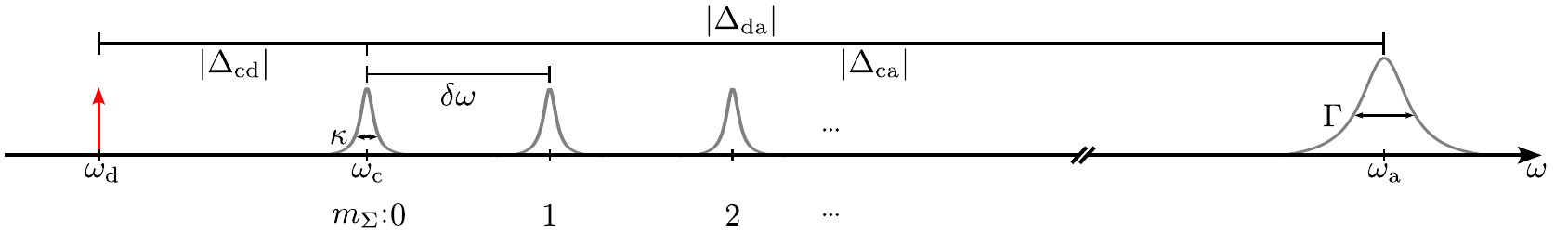} 
    \caption{\textbf{Sketch of frequencies.}
        Location in frequency space (horizontal axis, not to scale) of the atomic ($\omega_\mathrm{a}$) and cavity resonance ($ \omega_\mathrm{c} $), with respective line-widths $\Gamma$ and $\kappa$.
        The multi-mode cavity has mode-families, labelled by $m_\Sigma$, at intervals $m_\Sigma  \delta \omega $ above the fundamental cavity resonance ($ \omega_\mathrm{c} $).
        The drive beam (red arrow) is red-detuned from the fundamental cavity mode by $\abs{ \Delta_{\mathrm{cd}} }$, which is in turn red-detuned from the atomic resonance by $\abs{ \Delta_{\mathrm{ca}} }$.
        The drive--atom detuning $ \Delta_{\mathrm{da}} $ is thus of magnitude $\abs{ \Delta_{\mathrm{da}} } = \abs{ \Delta_{\mathrm{ca}} } + \abs{ \Delta_{\mathrm{cd}} }$.        
    }
    \label{f:freqs}
\end{figure}
Here we discuss the energy scales relevant to our cQED proposal.
We envision the scenario depicted in Fig.~\ref{f:freqs}, where the drive beam is detuned by an amount $ \Delta_{\mathrm{cd}}  \equiv  \omega_\mathrm{c}  - \omega_\mathrm{d}$ from the fundamental (TEM$_{00}$) mode of the cavity at frequency $ \omega_\mathrm{c}  \equiv \omega_{m=0}$, which in turn is far detuned from the bare atomic frequency $\omega_\mathrm{a}$, by an amount $ \Delta_{\mathrm{ca}}  \equiv  \omega_\mathrm{c}  - \omega_\mathrm{a}$.
The drive--atom detuning is thus $ \Delta_{\mathrm{da}}  =  \Delta_{\mathrm{ca}}  -  \Delta_{\mathrm{cd}} $.
We assume red detunings, i.e., that $\omega_\mathrm{d} <  \omega_\mathrm{c}  < \omega_\mathrm{a}$, such that $\abs{ \Delta_{\mathrm{da}} } = \abs{ \Delta_{\mathrm{ca}} } + \abs{ \Delta_{\mathrm{cd}} }$.

The transverse cavity mode frequencies $ \omega_m $ are indicated in the sketch of Fig.~\ref{f:freqs} at regular intervals above the fundamental cavity mode.
The size of the intervals is given by the transverse-mode-spacing (TMS) $ \delta \omega $, such that $ \omega_m  =  \omega_\mathrm{c}  + (n_x + n_y)  \delta \omega $, where $n_x,n_y$ are the transverse-mode indices of the $m$th mode, which is contained in the $m_\Sigma$th mode-family ($m_\Sigma\equiv n_x + n_y$).
The light--matter coupling of the $m$th cavity mode $ \Omega_m  \propto \sqrt{ \omega_\mathrm{c}  + (n_x + n_y) \delta \omega }$ can be approximated as $  \Omega_m   \approx \Omega_{m=0}$:
Whilst $ \delta \omega $ depends on the cavity's deviation from confocality, for optical cavities $ \omega_\mathrm{c}  / 2 \pi$ is on the order of hundreds of THz, such that
one can reasonably assume $ \delta \omega / \omega_\mathrm{c}  \ll 1$ \cite{Vaidya_etal2018, Sauerwein_etal2022}.
The cavity loss rate is parameterised by $\kappa$.

The frequency of the harmonic trap along the cavity axis is assumed to be large enough to produce a quasi-two-dimensional fermionic cloud, extended radially, i.e., transverse to the cavity axis.
The relevant trapping frequency is thus that along the radial direction, which we denote as $\omega_\mathrm{trap}$.

The loss rate due to spontaneous emission of the atoms is parameterised by $\Gamma$.
In the subsequent derivations, we will work in the dispersive regime $\abs{\Gamma /  \Delta_{\mathrm{da}} } \ll 1$, to allow for the adiabatic elimination of the excited internal atomic state.
This is achieved by far-detuning the cavity from the bare atomic transition such that $ \Delta_{\mathrm{ca}} $ is the dominant energy scale in the system.
In so doing, care must be taken to account for the non-zero TMS, which imposes a lower bound on $ \Delta_{\mathrm{ca}} $, depending on the number of TCMs mediating the atom--atom interactions.

A reasonable hierarchy of the above parameters in cQED platforms \cite{Vaidya_etal2018, Roux_etal2020, Zhang_etal2021, Roux_etal2021, Sauerwein_etal2022} is given by
\begin{equation}\label{e:hierarchy}
    \Delta_{\mathrm{ca}}  \gg  \Delta_{\mathrm{cd}} ,  \delta \omega  \gg \Gamma >  \Omega_m  > \kappa \gg \omega_\mathrm{trap} ,
\end{equation}
where the Rabi frequency of the drive $\Omega_{\mathrm{d} }$ can be tuned across this hierarchy, with feasible strengths ranging from $0$ to the order of several $\si{\GHz}$.

We will neglect atomic and cavity losses throughout our derivation, as they can be treated as subdominant in the dispersive regime:
already for $ \Delta_{\mathrm{da}}  / 2\pi \sim \SI{1}{\GHz}$, photonic and atomic losses such as those reported in \cite{Sauerwein_etal2022} occur at timescales suppressed by $\abs{\kappa /  \Delta_{\mathrm{da}} } \sim 10^{-4}$, $\abs{\Gamma /  \Delta_{\mathrm{da}} } \sim 10^{-3}$. Eventually, like for all methods based on cavity QED, the finite cooperativity of the cavity will limit the total duration available for the coherent evolution.

The energy scale of the speckled AC-Stark shift is discussed in the next section.

\subsection{Spatially dependent light shift}\label{s:acStark}
In this section, we describe how the interaction amplitudes $\mathcal{J}_{i_1 i_2 ; j_1 j_2} $ summarised by Eqs.~\eqref{e:Jmnpq}--\eqref{e:Jmnpq_asym} can be randomised, as required by the SYK model.
The key idea is to subject the cloud of fermionic atoms to a spatially disordered AC-Stark shift, which then translates into the effective model as a random contribution to the interaction integral of Eq.~\eqref{e:interaction_integral}.
Specifically, we envision the scenario utilized in Ref.~\cite{Sauerwein_etal2022}, where a light-shifting beam is used to off-resonantly dress the excited-state $\ket{e}$ with an auxiliary-state $\ket{a}$ (see also Fig.~\ref{f:concept}b of the main text).
The excited-state is then shifted in energy by an amount $\abs{ \Omega_{\mathrm{b}}( \boldsymbol{r} ) }^2 / (4 \Delta_\mathrm{b}) $ when the light-shifting beam is far-detuned from the excited-to-auxiliary state transition by an amount $\abs{\Delta_\mathrm{b}}$, where $\Delta_\mathrm{b} > 0$ for blue detuning, and $\Delta_\mathrm{b} < 0$ for red detuning.
Any spatial variation of the beam's Rabi frequency $\Omega_{\mathrm{b}}( \boldsymbol{r} )$ thus introduces a spatial dependence of the atoms' excited-state energy, and hence also of the drive--atom detuning
\begin{equation}\label{e:lightshift}
    \Delta_{\mathrm{da}} ( \boldsymbol{r} ) \equiv \omega_\mathrm{d} - \omega_\mathrm{a}( \boldsymbol{r} ) = (\omega_\mathrm{d} - \omega_\mathrm{a}) - \frac{\abs{ \Omega_{\mathrm{b}}( \boldsymbol{r} ) }^2 }{ 4 \Delta_\mathrm{b} } = \begin{cases}
        -\left( \abs{ \Delta_{\mathrm{da}} } + \frac{ \abs{ \Omega_{\mathrm{b}}( \boldsymbol{r} ) }^2 }{ 4 \abs{\Delta_\mathrm{b}} } \right) & \text{for } \Delta_\mathrm{b} > 0 , \\
        -\left( \abs{ \Delta_{\mathrm{da}} } - \frac{ \abs{ \Omega_{\mathrm{b}}( \boldsymbol{r} ) }^2 }{ 4 \abs{\Delta_\mathrm{b}} } \right) & \text{for } \Delta_\mathrm{b} < 0.
    \end{cases} 
\end{equation}
Here we have used that $ \Delta_{\mathrm{da}}  < 0$, i.e. that the drive frequency $\omega_\mathrm{d}$ is red detuned from the ground-to-excited state transition.
We choose the light-shifting beam to be blue detuned, such that the light shift increases the magnitude of $ \Delta_{\mathrm{da}} ( \boldsymbol{r} )$.
The above shows that a spatial disorder of $ \Delta_{\mathrm{da}} ( \boldsymbol{r} )$ can be engineered via a spatial disorder of the light-shifting beam's intensity $ I_{\mathrm{b}}( \boldsymbol{r} ) \propto \abs{\Omega_{\mathrm{b}}( \boldsymbol{r} ) }^2·$, and this translates into an effective model with random two-body interactions $\mathcal{J}_{i_1 i_2 ; j_1 j_2}$ by randomising the spatial integral of Eq.~\eqref{e:interaction_integral}.

To produce such a random intensity distribution, we propose to utilise an optical speckle (see for instance Ref.~\cite{Goodman2006} or the thesis~\cite{Jendrzejewski_PhDthesis}).
This can, for instance, be achieved by letting the beam pass through a diffuser or, in a more reproducible manner, by utilizing a spatial-light-modulator, and then focusing the beam into the atomic cloud with a lens \cite{Sauerwein_etal2022}.
The speckle pattern is characterised by grains of high intensity, randomly distributed in space, whose correlation length across the lens' focal plane is set by the wavelength of the light-shift beam, and the numerical aperture of the lens (Rayleigh criterium).
The distribution of the light intensity $I$ of the speckle pattern follows an exponential probability distribution $P(I) = \exp(-I/\langle I \rangle) / \langle I \rangle$ with $\langle I \rangle $ being the mean intensity (proportional to that of the light-shift beam) and a measure of the strength of the disorder.
Tuning the laser's intensity thus allows one to tune between weak and strong disorder.

\subsection{Including motional degrees of freedom}\label{s:Hkt_in_derivation}
Here, we demonstrate that the dynamics due to the motional (external) atomic degrees of freedom
\begin{equation}
    H_{\mathrm{kt}}  = \sum_{\mathrm{s}=\mathrm{e},\mathrm{g}} \int d^2r  \psi_\mathrm{s}^\dagger( \boldsymbol{r} ) \bigl( \boldsymbol{p}^2 / (2  m_{\mathrm{at}}  ) +  V_{\mathrm{t}} ( \boldsymbol{r} ) \bigr)  \psi_\mathrm{s} ( \boldsymbol{r} ) 
\end{equation}
only modify the one-body Hamiltonian of Eq.~\eqref{e:H_1body}, and thus its eigenmodes, but do not formally alter the effective two-body interaction of Eqs.~\eqref{e:Jmnpq}--\eqref{e:Jmnpq_asym}.
To do so, we consider the role of $ H_{\mathrm{kt}} $ in the three transformations of the original many-body Hamiltonian, outlined in Methods Sec.~\ref{s:derivation_Heff}:
\begin{enumerate}[(i)]
    \item \emph{Rotating frame} \\
    $ H_{\mathrm{kt}} $ is unaltered by the rotating frame transformation generated by $ H_\mathrm{RF} $ [defined above Eq.~\eqref{e:H_tindep}], as is evident from $\comm{  H_{\mathrm{kt}}  }{  H_\mathrm{RF}  }=0$. 
    Intuitively, this should be so, since $ H_\mathrm{RF} $ is proportional to the total number operator for the excited states, and $ H_{\mathrm{kt}} $ conserves particle number.
    For completeness, we show explicitly that this is indeed true. 
    To this end, we decompose the field operators $ \psi_\mathrm{g} ( \boldsymbol{r} )$, $ \psi_\mathrm{e} ( \boldsymbol{r} )$, respectively, into an arbitrary basis of mode-functions
    \begin{equation}\label{e:psi_decomp}
        \psi_\mathrm{s} ( \boldsymbol{r} ) = \sum_i \phi_{\mathrm{s} i}( \boldsymbol{r} ) c_{\mathrm{s} i} .
    \end{equation}
    In terms of this decomposition, $ H_{\mathrm{kt}} $ is given by 
    \begin{equation}\label{e:Hkt_decomp}
        H_{\mathrm{kt}}  = \sum_{\mathrm{s}=\mathrm{e},\mathrm{g}}   \sum_{i,j} \left( \int d^2r \phi_{\mathrm{s} i}^*( \boldsymbol{r} ) \left(  -\frac{\nabla^2_{ \boldsymbol{r} }}{2  m_{\mathrm{at}} }  +  V_{\mathrm{t}} ( \boldsymbol{r} ) \right) \phi_{\mathrm{s} j}( \boldsymbol{r} ) \right) c^\dagger_{\mathrm{s} i} c_{\mathrm{s} j}.
    \end{equation}
    This decomposition allows us to decouple the Laplacian in $ H_{\mathrm{kt}} $ from the operator algebra, which simplifies the calculation of commutators here and below. For the present case, we find
    \begin{equation}
        \comm{  H_{\mathrm{kt}}  }{  H_\mathrm{RF}  } \propto \comm{ c^\dagger_{\mathrm{s} i} c_{\mathrm{s} j} }{ \sum_k c^\dagger_{\mathrm{e} k} c_{\mathrm{e} k} } ,
    \end{equation}
    where we have used that the fermionic part of $ H_\mathrm{RF} $ is proportional to $\sum_k c^\dagger_{\mathrm{e} k} c_{\mathrm{e} k}$.
    The commutator on the right-hand-side vanishes since the operator on its left conserves the total (excited-state) particle number symmetry encoded by the operator on its right.
    So we see that $ H_{\mathrm{kt}} $ propagates unaltered into the rotating-frame Hamiltonian given by Eq.~\eqref{e:H_tindep}, i.e., Eq.~\eqref{e:H_tindep} changes to $H +  H_{\mathrm{kt}} $.
    
    \item \emph{Adiabatic elimination} \\
    The inclusion of $ H_{\mathrm{kt}} $ in Eq.~\eqref{e:H_tindep} modifies the Heisenberg equation of motion for $ \psi_\mathrm{e} ( \boldsymbol{r} )$ due to the additional commutator
    \begin{equation}
        \comm{ \psi_\mathrm{e} ( \boldsymbol{r} )}{ H_{\mathrm{kt}} } =  \left( \frac{\boldsymbol{p}^2}{2  m_{\mathrm{at}} } +  V_{\mathrm{t}} ( \boldsymbol{r} ) \right)  \psi_\mathrm{e} ( \boldsymbol{r} ) ,
    \end{equation}
    which can be calculated, as before, using the decomposition of Eq.~\eqref{e:Hkt_decomp}. The modified equation of motion is thus
    \begin{equation}
        i \partial_t  \psi_\mathrm{e} ( \boldsymbol{r} ) = - \Delta_{\mathrm{da}} ( \boldsymbol{r} )  \psi_\mathrm{e} ( \boldsymbol{r} ) + \Phi( \boldsymbol{r} )  \psi_\mathrm{g} ( \boldsymbol{r} ) +  \left( \frac{\boldsymbol{p}^2}{2  m_{\mathrm{at}} } +  V_{\mathrm{t}} ( \boldsymbol{r} ) \right)  \psi_\mathrm{e} ( \boldsymbol{r} ) .
    \end{equation}
    We drop the term in parenthesis, which is motivated by a separation of energy scales \cite{Maschler_etal2008};
    Having already assumed $ \Delta_{\mathrm{da}} ( \boldsymbol{r} )$ to be the dominant energy scale in the system at all $ \boldsymbol{r} $, the comparatively slow dynamics of the atoms' external degrees of freedom (typically on the order of tens of $\SI{}{\kHz}$ \cite{Roux_etal2021}) may be safely neglected.
    
    The steady-state field operator given by Eq.~\eqref{e:psi_e_adelim_methods} thus remains unchanged within this approximation, so that the Hamiltonian of Eq.~\eqref{e:H_adelim} simply acquires the additional term
    \begin{equation}
        \int d^2r  \psi_\mathrm{g}^\dagger( \boldsymbol{r} ) \left( \frac{\boldsymbol{p}^2}{2  m_{\mathrm{at}} } +  V_{\mathrm{t}} ( \boldsymbol{r} ) \right)  \psi_\mathrm{g} ( \boldsymbol{r} ) + \int d^2r \frac{\Phi^\dagger( \boldsymbol{r} ) \psi^\dagger_g( \boldsymbol{r} )}{ \Delta_{\mathrm{da}} ( \boldsymbol{r} ) } \left( \frac{\boldsymbol{p}^2}{2  m_{\mathrm{at}} } +  V_{\mathrm{t}} ( \boldsymbol{r} ) \right) \frac{\Phi( \boldsymbol{r} )  \psi_\mathrm{g} ( \boldsymbol{r} )}{ \Delta_{\mathrm{da}} ( \boldsymbol{r} ) } ,
    \end{equation}
    the latter part of which can be dropped since it is $\mathcal{O}(1/ \Delta_{\mathrm{da}} )$ relative to the Hamiltonian of Eq.~\eqref{e:H_adelim}.
    So, taking $ H_{\mathrm{kt}} $ into account during the adiabatic elimination of $ \psi_\mathrm{e} ( \boldsymbol{r} )$,  simply adds to Eq.~\eqref{e:H_adelim} the dynamics of the external degrees of freedom of the ground state species $ H_{\mathrm{kt}}  = \int d^2r \psi^\dagger( \boldsymbol{r} ) \left( \boldsymbol{p}^2 / (2  m_{\mathrm{at}} ) +  V_{\mathrm{t}} ( \boldsymbol{r} ) \right) \psi( \boldsymbol{r} )$, where we have denoted $ \psi_\mathrm{g} ( \boldsymbol{r} )$ as $\psi( \boldsymbol{r} ) = \sum_i \phi_i( \boldsymbol{r} ) c_i$ and redefined $ H_{\mathrm{kt}} $ accordingly as
    \begin{equation}\label{e:Hkt_post_ae}
        H_{\mathrm{kt}}  = \sum_{i,j} \left( \int d^2r \phi_{i}^*( \boldsymbol{r} ) \left(  -\frac{\nabla^2_{ \boldsymbol{r} }}{2  m_{\mathrm{at}} }  +  V_{\mathrm{t}} ( \boldsymbol{r} ) \right) \phi_{j}( \boldsymbol{r} ) \right) c^\dagger_{i} c_{j}.
    \end{equation}
    
    \item \emph{Schrieffer--Wolff transformation} \label{item:Hkt_in_SWT} \\
    The presence of $ H_{\mathrm{kt}} $ in the Hamiltonian obtained after adiabatic elimination merely modifies $H_0$ of Eq.~\eqref{e:SWT_H0} to $H_0 +  H_{\mathrm{kt}} $, i.e., it does not couple photonic and atomic degrees of freedom. 
    We may therefore continue to use the generator $S$ of Eq.~\eqref{e:SWT_generator_methods} to eliminate the coupling term $V$ given by Eq.~\eqref{e:SWT_V}. 
    All that remains is to take into account the additional contributions to the commutator $\comm{S}{H_0}$, which is modified to $\comm{S}{H_0} + \comm{S}{ H_{\mathrm{kt}} }$.
    We now show that this additional term vanishes, thereby proving that the effective Hamiltonian of Eq.~\eqref{e:H_post_SWT} only changes from $ H_\mathrm{eff} $ to $ H_\mathrm{eff}  +  H_{\mathrm{kt}} $.
    
    To simplify the notation, we start by summarising Eq.~\eqref{e:Hkt_post_ae} as ${ H_{\mathrm{kt}}  = \sum_{i,j}K(i,j)  c^\dagger_{i} c_{j}}$.
    Similarly, we group all scalar terms in Eq.~\eqref{e:SWT_generator_methods} such that ${ S= \sum_m  \sum_{k,l} s(m,k,l)  a_m  c_k^\dagger c_l - \mathrm{H.c.} }$.
    With this in hand, we have
    \begin{equation}
        \comm{S}{ H_{\mathrm{kt}} } =  \sum_m   a_m  \sum_{j,k} \left[ \sum_{i} \left( K(i,j) s(m,k,i)  - K(i,k)^* s(m,i,j) \right) \right] c^\dagger_k c_j + \mathrm{H.c.} \, .
    \end{equation}
    The sum over $i$ is $\sum_{i} \phi_{i}( \boldsymbol{r} )^* \phi_{i}( \boldsymbol{r} ') = \delta( \boldsymbol{r}  -  \boldsymbol{r} ')$, and so the term in square-brackets is proportional to
    \begin{equation}
        \int d^2r \frac{ g_\mathrm{d}^*( \boldsymbol{r} )  g_m ( \boldsymbol{r} )}{  \Delta_{\mathrm{da}} ( \boldsymbol{r} ) } \left\lbrace  \phi_k^*( \boldsymbol{r} ) \left(  -\frac{\nabla^2_{ \boldsymbol{r} }}{2  m_{\mathrm{at}} }  +  V_{\mathrm{t}} ( \boldsymbol{r} ) \right) \phi_{j}( \boldsymbol{r} ) -   \left[ \left(  -\frac{\nabla^2_{ \boldsymbol{r} }}{2  m_{\mathrm{at}} }  +  V_{\mathrm{t}} ( \boldsymbol{r} ) \right) \phi_{k}( \boldsymbol{r} ) \right]^* \phi_j( \boldsymbol{r} ) \right\rbrace .
    \end{equation}
    Multiplying this by $c^\dagger_k c_j$ and summing over $k,j$ reduces the term in curly braces above to 
    \begin{equation}
        \psi^\dagger( \boldsymbol{r} ) \left(  -\frac{\nabla^2_{ \boldsymbol{r} }}{2  m_{\mathrm{at}} }  +  V_{\mathrm{t}} ( \boldsymbol{r} ) \right) \psi( \boldsymbol{r} ) - \mathrm{H.c.} \, .
    \end{equation}
    This is zero, by hermiticity of $ H_{\mathrm{kt}} $, and so $\comm{S}{ H_{\mathrm{kt}} } = 0$.    
    Therefore, the effective Hamiltonian of Eq.~\eqref{e:H_post_SWT} is simply modified to
    $ H_\mathrm{eff}  = H_0 +  H_{\mathrm{kt}}  -  \sum_m  \frac{\Theta_m^\dagger \Theta_m}{  \Delta_m } $. This merely redefines the eigenmodes and eigen-energies of the one-body term in Eq.~\eqref{e:Heff}, but otherwise leaves the formal results of Eqs.~\eqref{e:Jmnpq}--\eqref{e:Jmnpq_asym} unchanged.
\end{enumerate}
In summary, we have shown that in the derivation of the effective model given by Eq.~\eqref{e:H_post_SWT}, one may disregard $ H_{\mathrm{kt}} $ when performing all necessary transformations, and simply add it back into the final one-body contribution in Eq.~\eqref{e:H_1body}.

\subsection{Compensating the disordered dipole potential}\label{s:compensation}
We have seen in the derivation of $ H_\mathrm{eff} $ [Eq.~\eqref{e:Heff}], that the one-body term [Eq.~\eqref{e:H_1body}] contains an effective disordered dipole potential
\begin{equation}\label{e:dipole_one}
    \int d^2r \frac{  \abs{\Omega_\mathrm{d}  g_\mathrm{d} ( \boldsymbol{r} )}^2  }{ \Delta_{\mathrm{da}} ( \boldsymbol{r} )} \psi^\dagger( \boldsymbol{r} ) \psi( \boldsymbol{r} ) ,
\end{equation}
which appears after adiabatic elimination of the excited state [see Eq.~\eqref{e:SWT_H0}].
Here we briefly discuss how this term could be compensated by introducing an additional drive.

The light-shift technique, described in Sec.~\ref{s:acStark}, produces in the dressed-state picture two energetically shifted states (Autler--Townes doublet).
The energy shifts of these dressed states, relative to the bare states, are perfectly anti-correlated.
In the microscopic Hamiltonian of Eq.~\eqref{e:sm_Hmb}, it is the energetically lower state of this doublet which is designated as the excited state, at frequency  $\omega_\mathrm{a}( \boldsymbol{r} )$.
Let the transition frequency from the ground state to the higher lying dressed state be $\omega_\mathrm{aux} ( \boldsymbol{r} )$, and consider an additional drive beam, at angular(Rabi) frequency $\omega_\mathrm{d'}$($\Omega_\mathrm{d'}$), detuned by $\Delta_\mathrm{d'aux}$ from this transition.
The microscopic model of Eq.~\eqref{e:sm_Hmb} is then modified by the additional terms
\begin{equation}
    \int d^2r \frac{\omega_\mathrm{aux} ( \boldsymbol{r} )}{2} \psi_\mathrm{aux}^\dagger ( \boldsymbol{r} ) \psi_\mathrm{aux} ( \boldsymbol{r} ) + 
    \Omega_\mathrm{d'} \int d^2r g_\mathrm{d'}( \boldsymbol{r} ) e^{-i\omega_\mathrm{d'} t} \psi_\mathrm{aux}^\dagger( \boldsymbol{r} )  \psi_\mathrm{g} ( \boldsymbol{r} )  + \mathrm{H.c.} \,.
\end{equation}

Going into the rotating frame generated by $ H_\mathrm{RF}  = \omega_\mathrm{d} \int d^2r  \psi_\mathrm{e}^\dagger ( \boldsymbol{r} )  \psi_\mathrm{e}  ( \boldsymbol{r} ) + \omega_\mathrm{d}  \sum_m  a^\dagger_m  a_m  + \omega_\mathrm{d'} \int d^2r \psi_\mathrm{aux}^\dagger ( \boldsymbol{r} ) \psi_\mathrm{aux} ( \boldsymbol{r} ) $, and adiabatically eliminating $ \psi_\mathrm{e} ( \boldsymbol{r} )$ and $\psi_\mathrm{aux}( \boldsymbol{r} )$, then yields the expression of Eq.~\eqref{e:SWT_H0}, but with an additional term 
\begin{equation}\label{e:dipole_two}
    \int d^2r \frac{ \abs{\Omega_\mathrm{d'} g_\mathrm{d'}( \boldsymbol{r} )}^2  }{\Delta_\mathrm{d'aux}( \boldsymbol{r} )} \psi^\dagger( \boldsymbol{r} ) \psi( \boldsymbol{r} ) .
\end{equation}
By tailoring the two drives such that they have matching intensity $\Omega_\mathrm{d}=\Omega_\mathrm{d'}$ and profiles $ g_\mathrm{d} ( \boldsymbol{r} ) =  g_\mathrm{d'}( \boldsymbol{r} )$, the two dipole terms [Eqs.~\eqref{e:dipole_one}~and~\eqref{e:dipole_two}] can be made to cancel by choosing their angular frequencies so as to achieve perfect anticorrelation of the respective detunings $ \Delta_{\mathrm{da}} ( \boldsymbol{r} ) = -\Delta_\mathrm{d'aux}( \boldsymbol{r} )$.
This can be achieved by choosing the detunings from the bare states to have equal magnitude, but opposite sign, since at any given position $ \boldsymbol{r} $ the dressed states' energy shifts are perfectly anti-correlated.

\subsection{A note on dissipation}\label{s:dissipation}
Here, we consider the role of losses in the effective model derived in Sec.~\ref{s:derivation_Heff}.
They arise due to spontaneous emission, at rate $\Gamma$, of the atomic excited state, and from out-coupling of the cavity modes, at rates $\kappa_m$.
For this open quantum system, we model the equation of motion of a given Heisenberg operator $O(t)$ via the adjoint master equation, which (for time-independent Lindblad generators) is \cite{BreuerPetruccioneTextBook},
\begin{equation}\label{e:adjoint_master_equation}
    \partial_t O(t) = i \comm{H_\mathrm{mb}}{O(t)} + \int d^2 r \left( L^\dagger( \boldsymbol{r} ) O(t) L( \boldsymbol{r} ) - \frac{1}{2} \acomm{L^\dagger( \boldsymbol{r} ) L( \boldsymbol{r} )}{O(t)} \right) +  \sum_m  \left( L_m^\dagger O(t) L_m - \frac{1}{2} \acomm{L_m^\dagger L_m}{O(t)} \right) ,
\end{equation}
where spontaneous emission of the atoms, and photon loss are, respectively, described by the jump operators $L( \boldsymbol{r} )=\sqrt{\Gamma} \psi_\mathrm{g}^\dagger( \boldsymbol{r} ) \psi_\mathrm{e} ( \boldsymbol{r} )$ and $L_m=\sqrt{\kappa_m} a_m $.
Here we have neglected the effect of atomic recoil due to spontaneous emission, which is equivalent to working at zeroth order in the Lamb--Dicke parameter $\eta$ \cite{GardinerZoller_Book2}. 
The first correction is of order $\eta^2$, and describes diffusion of the atoms due to spontaneous emission.
Here, we focus on the dynamics of the atoms' internal degrees-of-freedom.

The equation of motion for $\psi_\mathrm{g}^\dagger ( \boldsymbol{r} ) \psi_\mathrm{e} ( \boldsymbol{r} )$ under the dynamics described by Eq.~\eqref{e:adjoint_master_equation} is 
\begin{equation}
    \partial_t  \left( \psi_\mathrm{g}^\dagger ( \boldsymbol{r} ) \psi_\mathrm{e} ( \boldsymbol{r} ) \right) = i \left( \Delta_{\mathrm{da}} ( \boldsymbol{r} ) + i \Gamma / 2 \right) \psi_\mathrm{g}^\dagger ( \boldsymbol{r} ) \psi_\mathrm{e} ( \boldsymbol{r} )  - i \Phi( \boldsymbol{r} ) \psi^\dagger_\mathrm{g} ( \boldsymbol{r} )  \psi_\mathrm{g} ( \boldsymbol{r} ) - i \Phi( \boldsymbol{r} ) \psi^\dagger_\mathrm{e} ( \boldsymbol{r} )  \psi_\mathrm{e} ( \boldsymbol{r} ) .
\end{equation}
Adiabatically eliminating $\psi_\mathrm{g}^\dagger ( \boldsymbol{r} ) \psi_\mathrm{e} ( \boldsymbol{r} )$, we obtain
\begin{equation}\label{e:psi_ge_ss}
    \psi_\mathrm{g}^\dagger ( \boldsymbol{r} ) \psi_\mathrm{e} ( \boldsymbol{r} ) = \frac{ \Phi( \boldsymbol{r} ) \psi^\dagger_\mathrm{g} ( \boldsymbol{r} )  \psi_\mathrm{g} ( \boldsymbol{r} ) }{ \Delta_{\mathrm{da}} ( \boldsymbol{r} ) + i \Gamma / 2 } ,
\end{equation}
where we have assumed the contribution from the $\psi^\dagger_\mathrm{e} ( \boldsymbol{r} )  \psi_\mathrm{e} ( \boldsymbol{r} )$ term to be sub-leading, since in the dispersive regime $\abs{\Omega_{\mathrm{d}}/ \Delta_{\mathrm{da}} } \ll 1$, the low-saturation limit is satisfied \cite{MetcalfcanDerStraten2002}.

Similarly, adiabatic elimination of the photonic operator $ a_m $ yields
\begin{equation}\label{e:cav_op_ss}
    a_m  \approx \frac{ - \Omega_\mathrm{d}   \Omega_m^*}{2( \Delta_m  - i \kappa_m / 2)} \int d^2 r  \frac{ g_\mathrm{d} ( \boldsymbol{r} )  g_m^*( \boldsymbol{r} ) }{ \Delta_{\mathrm{da}} ( \boldsymbol{r} )  + i\Gamma/2 } \psi^\dagger_\mathrm{g} ( \boldsymbol{r} )  \psi_\mathrm{g} ( \boldsymbol{r} ) ,
\end{equation}
where we dropped terms of order $\Omega_m / \Omega_{\mathrm{d}}$, motivated by the discussion of Sec.~\ref{s:energies}.

Inserting the operators of Eq.~\eqref{e:psi_ge_ss}~and~\eqref{e:cav_op_ss} into Eq.~\eqref{e:adjoint_master_equation} then yields an effective dissipator 
\begin{equation}
    \mathcal{D}_{\mathrm{eff}} \bullet  = \int d^2 r \left( L_{\mathrm{eff}}( \boldsymbol{r} ) \bullet L_{\mathrm{eff}}^\dagger( \boldsymbol{r} ) - \frac{1}{2} \acomm{L_{\mathrm{eff}}^\dagger( \boldsymbol{r} ) L_{\mathrm{eff}}( \boldsymbol{r} )}{\bullet} \right) +  \sum_m  \left( L_{m,\mathrm{eff}} \bullet L_{m,\mathrm{eff}}^\dagger - \frac{1}{2} \acomm{L_{m,\mathrm{eff}}^\dagger L_{m,\mathrm{eff}}}{\bullet} \right),
\end{equation}
describing dephasing of the remaining internal atomic degree-of-freedom (up to order $1/ \Delta_{\mathrm{da}}^2$ included), via the effective jump operators
\begin{equation}
    L_{\mathrm{eff}}( \boldsymbol{r} ) = \frac{\sqrt{\Gamma}}{\Delta_{\mathrm{da}} ( \boldsymbol{r} ) +i\Gamma/2} \Omega_\mathrm{d}  g_\mathrm{d} ( \boldsymbol{r} )  \psi_\mathrm{g}^\dagger( \boldsymbol{r} )  \psi_\mathrm{g} ( \boldsymbol{r} ),
\end{equation}
and
\begin{align}
    L_{m,\mathrm{eff}} =& \sqrt{\kappa_m} \frac{ \Omega_\mathrm{d}   \Omega_m^* }{2( \Delta_m - i\kappa_m/2 ) } \int d^2 r  \frac{ g_\mathrm{d} ( \boldsymbol{r} )  g_m^*( \boldsymbol{r} ) }{ \Delta_{\mathrm{da}} ( \boldsymbol{r} )  + i\Gamma/2 } \psi_\mathrm{g}^\dagger( \boldsymbol{r} )  \psi_\mathrm{g}( \boldsymbol{r} ) .
\end{align}
The integral over the atomic cloud's volume in $L_m^{\mathrm{(eff)}}$ reflects a ``global dephasing'' arising from the fact that a photon emitted via the cavity mirrors leaves the observer ignorant as to the position $ \boldsymbol{r} $ at which the photon was scattered by an atom.

The effective jump operators are randomised via the disordered detuning $ \Delta_{\mathrm{da}} ( \boldsymbol{r} )$, thus yielding random, quadratic jump operators.
This is similar to the dissipative SYK model studied in Ref.~\cite{Lucas_etal2022}.

\subsection{SYK model with Cauchy distribution}\label{s:cauchy_gauss}
Here, we discuss the spectral properties of the model in Eq.~\eqref{e:Hsyk} with the interaction amplitudes drawn from the Cauchy distribution, and compare them to that of the target model.
The ideal Cauchy distribution $P(x)$ is defined over the domain $x \in (-\infty, \infty)$, and the normalised distribution function centred at $x=0$ is given by
\begin{equation}\label{e:idealcauchy}
    P(x) = \frac{1}{\pi}\frac{\gamma} {\gamma^2 + x^2}, 
\end{equation}  
where $\gamma$ is the half-width at half-maximum of the distribution. 
Since the moments of this distribution are not defined, in our numerics we truncate the domain to $x \in [-a, a]$, which in turn results in a domain-dependent variance of the distribution.
We choose the value of $a$ such that we cover a desired fraction $f \equiv \int_{-a}^{a} dx P(x)=0.975$ of the probability mass of Eq.~\eqref{e:idealcauchy}.
With this rationale, one has the relation $a = \gamma\tan(f\pi/2)$, and the normalised truncated Cauchy distribution function is then given by
\begin{equation}
    \label{truncauchy}
    P_a(x) = \frac{1}{2\arctan(a/\gamma)} \frac{\gamma} {\gamma^2 + x^2},
\end{equation}
which can be tuned by changing the width $\gamma$ for a fixed $f$.

\begin{figure}[h!]
    \centering
    \includegraphics[width=.49\linewidth]{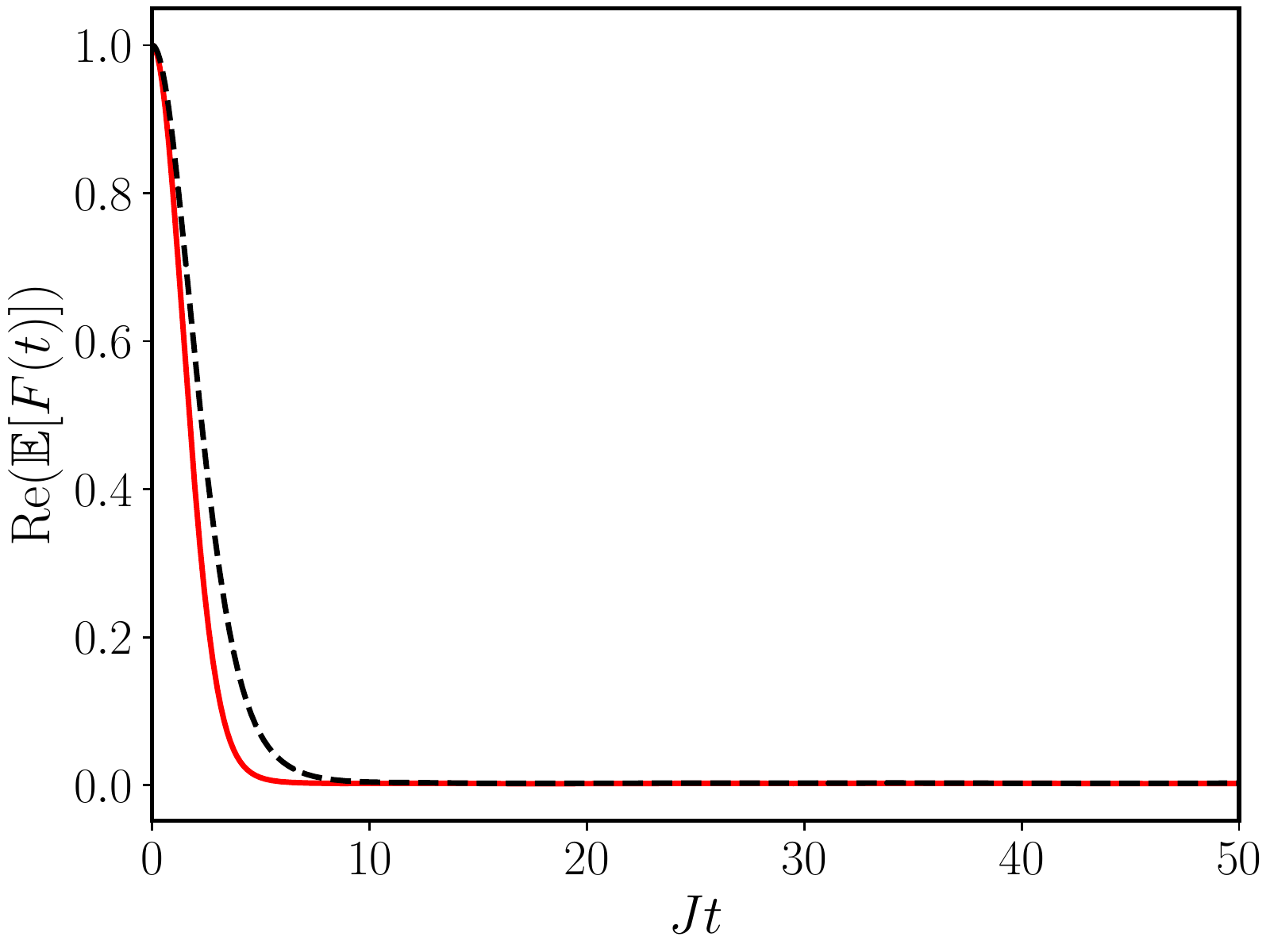}
    \includegraphics[width=.49\linewidth]{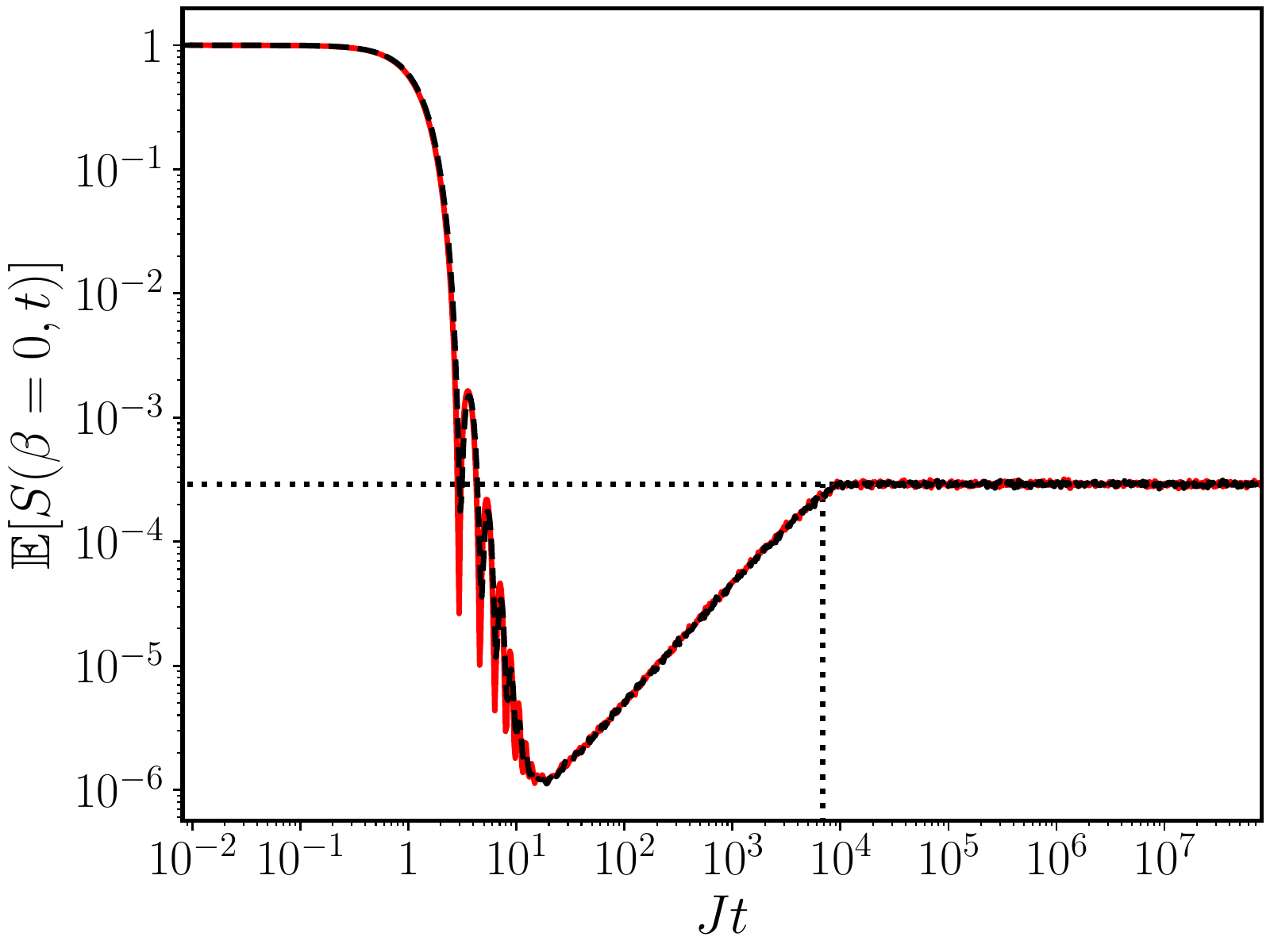}
    \caption{\textbf{Comparison of SYK model with Cauchy-distributed interactions.} 
        Comparison of OTOC (left) and SFF (right) dynamics, generated by the target Hamiltonian $H_\mathrm{SYK}$ with complex Gaussian(truncated Cauchy) distributed $ J_{i_1 i_2 ; j_1 j_2} $, as indicated by the red solid(black dashed) curves.
        The OTOC(SFF) data are for a system of $N=10$($N=14$) fermionic modes at half-filling, and averaged over $1000$ realizations.
        For the OTOCs, the choice of operators is the same as for the OTOCs shown in Fig.~\ref{f:otocs} of the main text.
        The time axes of the SFF curves were rescaled so as to match their respective Heisenberg times.
        Dotted black curves are as in Fig.~\ref{f:sff} of the main text.
    }
    \label{f:compare_GCExp}
\end{figure}

We draw interaction amplitudes $ J_{i_1 i_2 ; j_1 j_2} $ from the distribution $P_a(x)$, and construct the matrix representation of the SYK model in the same way as for the usual Gaussian definition [Eq.~\eqref{e:Hsyk}].
In Fig~\ref{f:compare_GCExp} we compare the OTOCs and the SFF generated by the target SYK model [Eq.~\eqref{e:Hsyk}] with complex Gaussian-distributed interactions (red solid curves) to the variation of the model with Cauchy-distributed interactions drawn from $P_a(x)$ for $\gamma=0.2$ (black dashed curves), averaged over $1000$ disorder realizations.
The OTOCs of the variant with Cauchy-distributed interactions decay slower than those of the model with Gaussian-distributed interactions.
For the SFF, the qualitative features (early-time power-law decaying oscillations, followed by a linear-in-$t$ ramp, and a plateau) agree well.
We note that in plotting the SFF, a slight difference in the Heisenberg times has been corrected for (by rescaling the time axes), which amounts to having done a spectral unfolding.

\subsection{On-axis drive}\label{s:lgtdl_drive}
As motivated in Sec.~\ref{s:derivation_Heff} of the Methods, and in the main text, one may consider an on-axis drive beam, which couples to the cavity modes instead of the atoms, in order to satisfy the long-wavelength approximation.
In this section, we show that this approach formally yields the same effective model as in Methods Sec.~\ref{s:derivation_Heff}, only with $ g_\mathrm{d} ( \boldsymbol{r} )$ replaced by a superposition of the cavity modes [see Eq.~\eqref{e:sup_camodes}].

Changing from a transverse to an on-axis drive, the many-body Hamiltonian of Eqs.~\eqref{e:sm_Hmb}~and~\eqref{e:H_tindep} is modified accordingly as
\begin{equation}\label{e:Hmb_lgtdl}
    \begin{split}
        H_\mathrm{mb} =& \sum_m  \Delta_m  a^\dagger_m  a_m  - \int d^2r  \Delta_{\mathrm{da}} ( \boldsymbol{r} )  \psi_\mathrm{e}^\dagger ( \boldsymbol{r} )  \psi_\mathrm{e}  ( \boldsymbol{r} ) + \frac{1}{2} \sum_m \int d^2r \left(  \Omega_m   g_m ( \boldsymbol{r} )   a_m   \psi_\mathrm{e}^\dagger( \boldsymbol{r} )  \psi_\mathrm{g} ( \boldsymbol{r} ) + \mathrm{H.c.} \right) \\
        &+  \sum_m (\Omega_\mathrm{d}^* c_m  a_m  + \mathrm{H.c.} ) ,
    \end{split}
\end{equation}
where now the last term represents the coupling of the on-axis drive to the transverse cavity modes, with coefficients $c_m$ quantifying the strength of this coupling to the $m$th cavity mode (i.e. the overlap integral of the relevant mode profiles at the cavity mirror).
As for the case of transverse drive, we neglect the kinetic and external trap terms $ H_{\mathrm{kt}} $ here and in what follows.
Note further that we have already moved into the frame rotating at the drive frequency $\omega_d$, and have included the spatial dependence of the atomic resonance $\omega_\mathrm{a}( \boldsymbol{r} )$ via the drive--atom detuning $ \Delta_{\mathrm{da}} ( \boldsymbol{r} )$.

Since the cavity modes are now driven, we decompose $ a_m $ into a sum of quantum fluctuations $\delta  a_m $ around a classical contribution $\alpha_m = \tr(\rho  a_m )$, $ a_m  = \alpha_m + \delta  a_m $.
The equation of motion for $ a_m $, including cavity losses $\kappa$, is
\begin{equation}
    i \partial_t  a_m  = (  \Delta_m  - i\kappa / 2)  a_m  + \Omega_d c_m^* + \int d^2r \left( \Omega_m g_m(r) \right)^*  \psi_\mathrm{g}^\dagger( \boldsymbol{r} )  \psi_\mathrm{e} ( \boldsymbol{r} ) .
\end{equation}
Taking the trace and equating zero order terms we obtain,
\begin{equation}
    i \partial_t \alpha_m = (  \Delta_m  - i\kappa / 2) \alpha_m + \Omega_d c_m^*,
\end{equation}
which is solved by $\alpha_m(t) = \alpha_m(0) \exp(-(i \Delta_m  +\kappa/2 )t ) - \Omega_{\mathrm{d}} c_m^* / ( \Delta_m  - i \kappa/2)$.
Assuming $\abs{ \Delta_m } \gg \kappa$, we find $\langle \alpha_m \rangle_T \equiv (1/T) \int_0^T dt \alpha_m(t) \simeq - \Omega_{\mathrm{d}} c_m^* /  \Delta_m $, where we have additionally assumed that the time interval satisfies $T \gg 2 \abs{\alpha_m(0) /  \Delta_m  }$, which is valid for sufficiently large cavity--drive detunings $ \Delta_m $.
Substituting $ a_m $ by  $\langle \alpha_m \rangle_T +  a_m $ in Eq.~\eqref{e:Hmb_lgtdl}, we obtain
\begin{equation}\label{e:H_lgtdl_mf}
    H_\mathrm{mb} =  \sum_m \left(  \Delta_m  a^\dagger_m  a_m  - \frac{\abs{\Omega_{\mathrm{d}} c_m}^2 }{ \Delta_m } \right) - \int d^2r  \Delta_{\mathrm{da}} ( \boldsymbol{r} )  \psi_\mathrm{e}^\dagger ( \boldsymbol{r} )  \psi_\mathrm{e}  ( \boldsymbol{r} )  + \int d^2r \left( \tilde{\Phi}( \boldsymbol{r} )  \psi_\mathrm{e}^\dagger( \boldsymbol{r} )  \psi_\mathrm{g} ( \boldsymbol{r} )  + \mathrm{H.c.} \right), 
\end{equation}
where now 
$\tilde{\Phi}( \boldsymbol{r} ) = \frac{1}{2} \sum_m \left( - \frac{\Omega_\mathrm{d} c_m^*}{ \Delta_m } +  a_m  \right) \Omega_m g_m( \boldsymbol{r} )$.
Comparing this with $\Phi( \boldsymbol{r} )$ of Eq.~\eqref{e:Phi} already indicates how the transverse drive profile $ g_\mathrm{d} ( \boldsymbol{r} )$ of Sec.~\eqref{s:derivation_Heff} is replaced here by a superposition of the cavity mode profiles, governed by the drive strength $ \Omega_\mathrm{d} $ and the coupling coefficients $c_m$.
In what follows, we drop the constant contained in the first term of $ H_\mathrm{mb}$.

Equation~\eqref{e:H_lgtdl_mf} is formally the same as Eq.~\eqref{e:H_tindep}, and thus adiabatic elimination of $ \psi_\mathrm{e} ( \boldsymbol{r} )$ and integration of the cavity photons $ a_m $ via the SWT proceed analogously to the derivation of Eq.~\eqref{e:H_post_SWT}.
Formally, we obtain the same result, only with modified terms,
\begin{align}
    H_0 &=  H_{\mathrm{kt}}  + \sum_m  \Delta_m  a^\dagger_m  a_m  + \frac{\abs{\Omega_{\mathrm{d}}}^2}{4} \int d^2r  \frac{1}{ \Delta_{\mathrm{da}} ( \boldsymbol{r} )} \sum_{m,n} \frac{ (c_m^* \Omega_m g_m( \boldsymbol{r} ) )^*  c_n^* \Omega_n g_n( \boldsymbol{r} ) }{\Delta_m \Delta_n} \psi^\dagger( \boldsymbol{r} ) \psi( \boldsymbol{r} ) , \\
    \Theta_m &=  -\frac{1}{4}\int d^2 r \Omega_{\mathrm{d}}^* \Omega_m g_m( \boldsymbol{r} ) \left( \sum_n \frac{c_n^* \Omega_n g_n( \boldsymbol{r} )}{\Delta_n} \right)^* \frac{\psi^\dagger( \boldsymbol{r} ) \psi( \boldsymbol{r} )}{ \Delta_{\mathrm{da}} ( \boldsymbol{r} )}.
\end{align}
The interactions 
\begin{equation}
    \mathcal{J}_{i_1 i_2 ; j_1 j_2} = \sum_{m} \frac{ I_{i_1 j_1, m} I_{j_2 i_2, m}^* }{ \Delta_m } = \mathcal{J}_{j_2 j_1 ; i_2 i_1}^* ,
\end{equation}
of the effective model with on-axis drive are thus formally equivalent to those of Eq.~\eqref{e:interaction_integral} with transverse drive, but with the interaction integrals of Eq.~\eqref{e:Jmnpq} modified to
\begin{equation}
    I_{i_1 j_1, m} = \int d^2r \frac{ \Omega_\mathrm{d} \left(  \Omega_m   g_m ( \boldsymbol{r} ) \right)^* \mu( \boldsymbol{r} ) \phi^*_{i_1}( \boldsymbol{r} ) \phi_{j_1}( \boldsymbol{r} ) }{4  \Delta_{\mathrm{da}} ( \boldsymbol{r} )} ,
\end{equation}
where
\begin{equation}\label{e:sup_camodes}
    \mu(r) = \sum_n \frac{c_n^* \Omega_n g_n( \boldsymbol{r} )}{\Delta_n} ,
\end{equation}
is the superposition of cavity modes to which the on-axis drive has non-zero coupling.
The coupling coefficients may be varied so as to achieve a desired profile $\mu(r)$.
For instance, one may engineer the on-axis drive so as to couple only to the lowest (Gaussian) cavity mode, $c_m=0$ for $m>0$, such that the drive profile $ g_\mathrm{d} ( \boldsymbol{r} )$ in Eqs.~\eqref{e:interaction_integral} is replaced by the Gaussian cavity mode.

\end{document}